\pdfoutput=1
\documentclass[epj]{svjour}
\usepackage{graphicx} 
\usepackage{amsmath} 
\usepackage{color}
\usepackage{calc}
\usepackage{booktabs}
\def\rvec {\mathbf{r}}

\newif\ifcomments

\commentsfalse

\begin{document}
\title{Structure of Mg$_n$ and Mg$^+_n$ clusters up to n\,=\,30}
\author{S. Janecek\inst{1,2} \and E. Krotscheck\inst{2,3} \and
  M. Liebrecht\inst{2} \and R. Wahl\inst{2,4} \thanks{This work was
    supported by the Austrian Science Fund FWF under projects P18134
    and P21924 (to EK) and J2936 (to SJ).  Computational support was
    provided by the Central Computing Services at the Johannes Kepler
    Universit\"at Linz, we would especially like to thank Johann
    Messner for help and advice in using the facility.}}

\institute{Institut de Ci\'{e}ncia de Materials de Barcelona
  (ICMAB--CSIC), Campus de Bellaterra, 08193 Barcelona,
  Spain\\
  \and Institute for Theoretical Physics,
  Johannes Kepler Universit\"at Linz, A-4040 Linz, Austria\\
  \and Department of Physics, University at Buffalo SUNY
  Buffalo NY 14260\\
  \and Faculty of Physics and Center for Computational 
  Materials Science, University of Vienna, A-1090
  Vienna, Austria }

\abstract{We present structure calculations of neutral and singly
  ionized Mg clusters of up to 30 atoms, as well as Na clusters of up
  to 10 atoms. The calculations have been performed using density
  functional theory (DFT) within the local (spin-)density
  approximation, ion cores are described by pseudopotentials. We have
  utilized a new algorithm for solving the Kohn-Sham equations that is
  formulated entirely in coordinate space and, thus, permits
  straightforward control of the spatial resolution. Our numerical
  method is particularly suitable for modern parallel computer
  architectures; we have thus been able to combine an unrestricted
  simulated annealing procedure with electronic structure calculations
  of high spatial resolution, corresponding to a plane-wave cutoff of
  954eV for Mg. We report the geometric structures of the resulting
  ground-state configurations and a few low-lying isomers. The
  energetics and HOMO-LUMO gaps of the ground-state configurations are
  carefully examined and related to their stability properties. No
  evidence for a non-metal to metal transition in neutral and
  positively charged Mg clusters is found in the regime of ion numbers
  examined here.}  
\maketitle

\section{Introduction}

The properties of nanometer-sized clusters of atoms are significantly
different from those of the isolated chemical species as well as the
bulk material, and generally exhibit a strongly non-monotonous
size-dependent behavior \cite{BrackPhysRep}. Clusters are the ultimate
nanostructures, where a fundamental understanding of their properties
can be achieved one atom and one electron at a time. They thus open a
unique window to study the emergence of the properties of macroscopic
matter from its microscopic constituents, and have become an active
field in basic research. Nanostructured materials also hold, due to
low dimensionality and unique composition, the promise of
technological applications reaching as far as clean and sustainable
energy, reactions, and catalysis \cite{Serrano:2009fv,Centi:2010dg}.

The transition from microscopic to macroscopic behavior is
particularly obvious for clusters of atoms that are bound by covalent
bonds at small atom numbers, but are metals in bulk quantities. These
must, at some point, undergo a transition to metallic behavior.

The history of the examination of these clusters is long; we take a
fresh view for a number of reasons: One is that we present here the
first large-scale application of \texttt{limerec} \cite{limerec}, a
new open source Density Functional Theory (DFT) package specifically
designed for cluster calculations. The method is formulated entirely
in real space. It thus avoids any basis set bias and is particularly
suited for modern parallel computing environments
\cite{cluster,nonloc,any}. Due to its efficiency, the method allows to
find ground state configurations and low-lying isomers of fairly large
clusters by unrestricted minimization of the total energy in the whole
configuration space, using Langevin-Monte-Carlo and/or steepest
descent annealing procedures. It thus also eliminates a possible bias
due to the choice of a certain symmetry or starting configuration that
is present in more restricted minimization schemes.  Our second
objective is to prepare for studying the influence of a quantum fluid
matrix on the formation and structure of metallic clusters in the near
future.

The main thrust of this paper is to present a detailed and systematic
characterization of neutral and singly ionized magnesium clusters of
up to 30 atoms. We also discuss some sample calculations for smaller
sodium clusters, mostly for the purpose of studying electron
localization.  All calculations have been performed in the framework
of spin-density functional theory (SDFT), using the Perdew-Wang
exchange-correlation functional \cite{PeW92} and local
\cite{PerdewPotentials,Kueb00} as well as norm-conserving non-local
pseudopotentials \cite{troullier1993} to treat the core electrons. In
addition to the spatial structures of the ground state configurations
and low-lying isomers, we also report binding energies, ionization
energies and fragmentation energies as well as the HOMO-LUMO gap for
all clusters.  Details of the computational method are given in
section \ref{sec:computing}.

We have chosen to study mostly Mg$_n$ and Mg$_n^+$ clusters for a
number of reasons: The simpler analog, Na$_n$ clusters, has been
studied extensively, see, for example, Ref.~\cite{IKAK09} for a recent
review as well as references to earlier work. Although there are still
some open issues \cite{deHeer2010}, Na$_n$ clusters are reasonably
well described by a simple jellium model, in particular energetic
quantities like magic numbers are well reproduced. In Mg$_n$ clusters,
electrons are more strongly localized and, hence, a jellium model is
less appropriate. Comparing Mg$_n$ and Na$_n$ clusters can therefore
tell us about the consequences of electron localization.

A striking feature is that Mg$_n$ clusters evolve, with increasing ion
number $n$, from molecule-like complexes bound by covalent bonds to a
bulk metal where the valence electrons are delocalized. For Mg
clusters in helium, this non-metal to metal (NMM) transition has been
reported to occur around $n=20$ with different experimental techniques
\cite{DDB01,diederichPRA05,Thomas02}. Thomas {\em et al.\/}
\cite{Thomas02}, for example, have measured photoelectron spectra of
mass-selected magnesium cluster anions, which are related to the
HOMO-(HOMO-1) gap. Similar results have been reported for Cd clusters
\cite{IssChe05}. We note, however, that the experimental evidence of
NMM transitions is rather
indirect. Refs.~\cite{Jellinek02,Acioli213402} argue that due to the
method of measurement, the observed NMM transition may even depend on the
formation process. For a review and discussion, see
Ref.~\cite{IssChe05}.

The size range where the NMM transition has been proposed is easily
accessible by DFT calculations.  We will contribute to the discussion
here by studying the development of HOMO-LUMO gaps as a function of
cluster size, and by looking at the degree of localization of the
electronic density.

Another objective of our work is to prepare for the study of metallic
clusters in quantum fluid matrices.  Techniques to agglomerate atoms
and small molecules in a quantum fluid matrix --- specifically, in
superfluid $^4$He --- have opened a new and versatile way to study the
structural, electronic and spectroscopic properties of
nanoparticles. The helium droplets can be viewed as ultracold
nanoscopic reactors, which isolate single molecules, clusters, or even
single reactive encounters at very low temperatures
\cite{toennies04}. Clusters of well-defined composition can be formed
inside the droplets, and their examination in the millikelvin regime
has already given important clues on magnetism and superconductivity
on the nanometer scale~\cite{Toennies98}.

There is increasing evidence that the presence of a helium matrix can
indeed change the geometric arrangement and other properties of the
metal cluster ions. For example, cluster growth by capturing in $^4$He
droplets can lead to different isomers, which are not found with other
cluster generation techniques and which may be affected differently by
the He environment. Such an effect has been observed for clusters
bound by hydrogen bridges: Cyclic water hexamers were found in helium
droplets, but not in vacuum~\cite{nautaScience00}. There is also
evidence that the feedback of the surrounding quantum fluid on the
much more strongly bound carbon nanotubes is non-negligible; Kim {\it
  et al.\/} \cite{ColeRelax} point out that this is basically dictated by
Newton's third law. An unambiguous interpretation of such experiments
requires understanding of the influence of the $^4$He surrounding the
cluster. Also, metallic clusters that are difficult to generate in
vacuum can be formed in a fluid matrix~\cite{ZC91}. Liquid helium
provides an ideal medium for such ``nanoreactors'' because it is
transparent in the entire spectral range from the far IR to vacuum UV,
and high spectroscopic resolution, comparable to the gas phase, can be
achieved.

Our paper is organized as follows: In Sec.~\ref{sec:computing} we give
a brief discussion of the computational methods used. The core of our
DFT package is a diffusion algorithm for solving Schr\"odinger-like
equations; a more extensive analysis of the method, including
convergence tests, a comparison with the implicitly restarted Lanczos
method and an assessment of its performance on parallel computing
architectures, has been given in Ref.~\cite{any}. A separate program
package for solving the Schr\"odinger equation for the bound states in
an arbitrary local potential is also available \cite{sch_anysolve}.
Sec.~\ref{sec:results} turns to the results of our calculations. We
briefly study Na$_n$ and Na$_n^+$ clusters and determine their
structure and energetics. A somewhat unexpected feature is that local
pseudopotentials predict a distorted ground state configuration of
Na$_4$. We then turn to our discussion of Mg$_n$ and Mg$_n^+$
clusters. We present results for the ground state configuration,
energetics, and stability of these systems and examine the appearance
and origin of ``magic numbers''.  The concluding section
\ref{sec:conclusion} gives a brief summary of our findings.

\section{Computational Methods}
\label{sec:computing}

\subsection{Spin-Density Functional Theory}

Spin-density functional theory \cite{PeW92} maps the solution of the
interacting many-electron problem onto that of a non-interacting
auxiliary system, which is described by the Kohn-Sham equations,
\begin{eqnarray}
  \left[ -\frac{\hbar^2}{2m} \nabla^2 +
    V^{\sigma}_\mathrm{KS} \left[\{ \rho^\sigma \} \right]
    \left(\mathbf{r}\right) \right]
    \psi^\sigma_j  \left(\mathbf{r}\right) &=& \epsilon^\sigma_j
    \psi^\sigma_j  \left(\mathbf{r}\right),
    \label{eq:ks1} \\
    \rho \left(\mathbf{r}\right) = \sum_\sigma \sum_j n_j^\sigma \left|
      \psi^\sigma_j  \left(\mathbf{r}\right) \right|^2 &\equiv& \sum_\sigma
    \rho^\sigma  \left(\mathbf{r}\right)\,.
      \label{eq:ks2}
\end{eqnarray}
These are non-linear Schr\"odinger equations for a set of
single-particle wave functions $\psi^\sigma_j \left(\rvec\right)$ in
an effective potential $V^{\sigma}_\mathrm{KS} \left[\{ \rho^\sigma \}
\right] \left(\mathbf{r}\right)$, where $\sigma$ denotes the spin
index and $n_j^\sigma$ is the occupation number of the state
$\{j,\sigma\}$. The effective potential
\begin{equation}
  \label{eq:vks}
  V^{\sigma}_\mathrm{KS} \left[\{ \rho^\sigma \} \right] \left(\mathbf{r}\right)
  = V_\mathrm{ext} \left(\mathbf{r},\left\{\mathbf{R}_i\right\}\right)
  + V_\mathrm{C} \left[\rho\right] \left(\rvec\right)
  + V_\mathrm{xc} \left[\left\{\rho^\sigma \right\}\right]
  \left(\mathbf{r}\right)
\end{equation}
consists of the external potential $V_\mathrm{ext}
\left(\mathbf{r},\left\{\mathbf{R}_i\right\}\right)$ describing the
interaction of the valence electrons with the ion cores located at the
positions $\{\mathbf{R}_i\}$, the Coulomb
term
\begin{equation}
V_\mathrm{C} \left[\rho\right] \left(\rvec\right) =\int
    \frac{e^2}{\left| 
        \mathbf{r}-\mathbf{r}' \right|}\rho \left(\mathbf{r}'\right)
    d^3\mathbf{r}'\,,
\end{equation}
which accounts for the direct electron-electron interaction, and the
exchange-correlation potential $V_\textrm{xc}[\{ \rho^\sigma
\}](\mathbf{r})$. In this work, the local spin-density approximation
\cite{vBH72} is used for $V_\mathrm{xc}$, with the correlation
functional proposed by Perdew and Wang \cite{PeW92}.

Equations (\ref{eq:ks1}) and (\ref{eq:ks2}) are solved
self-consistently for each set of ion positions $\{\mathbf{R}_i\}$. To
find the global minimum of the total energy, the energy function
$E(\{\mathbf{R}_i\})$ must be evaluated at many different points in
configuration space. Thus, an indispensable prerequisite for
performing an unrestricted minimization is a fast and reliable method
to solve the Kohn-Sham equations.

\subsection{Diffusion algorithm}
\label{ssec:diffusion}

In recent years, real-space methods for solving the Kohn-Sham equations
\eqref{eq:ks1},~\eqref{eq:ks2} have become increasingly popular
\cite{BeckDFT,torsti06}. They are easy to implement,
free from modeling uncertainties, and well suited for modern massively
parallel computers. The calculations in this work have been performed
using the program package \texttt{limerec} \cite{limerec}, which
solves the Kohn-Sham equations on a real-space grid using an efficient
diffusion algorithm: The lowest $n$ eigenstates of the one-body
Schr\"odinger equation
\begin{equation}
(T+V)\psi_j({\bf r}) = E_j\psi_j({\bf r})
\end{equation}
are obtained by repeatedly applying the imaginary time evolution operator
\begin{equation}
{\cal T}(\epsilon)  \equiv e^{-\epsilon(T+V)}
\label{evol}
\end{equation}
to a set of trial functions $\bigl\{\psi_j({\bf r})\bigr\}$, which are
orthogonalized after each step. Above, $T$ is the one-body kinetic
energy operator, and $V$ the potential. The operator ${\mathcal
  T}(\epsilon)$ multiplies states corresponding to higher energies by
exponentially decreasing weights. It thus gradually filters out
high-energy states, making the procedure converge to the lowest $n$
eigenstates of the operator $T+V$.

The evolution operator \eqref{evol} can not be calculated exactly.
A simple approximation is the so-called split operator form
\cite{feit2},
\begin{equation}
 {\cal T}_2(\epsilon)\equiv e^{-{1\over
    2}\epsilon V} e^{-\epsilon T} e^{-{1\over 2}\epsilon V} =
e^{-\epsilon\left(T+V +  \mathcal{O}(\epsilon^2)\right)}\,;
\label{so}
\end{equation}
it is accurate up to second order in the timestep $\epsilon$.
 
Factorizing $\mathcal{T}(\epsilon)$ obviously reduces the problem of
calculating the exponential of the full Hamiltonian to the problem of
dealing with its parts separately. For a local operator $V({\bf r})$,
the action of $e^{-\frac{1}{2}\epsilon V({\bf r})}$ on a function
$\psi_j({\bf r})$ is just a vector-vector multiplication. The operator
$e^{-\epsilon T}$ is local in momentum space, its action on a given
state needs one Fast Fourier Transform (FFT) pair (forward and
backward) in addition to a vector-vector multiplication in Fourier
space. If $\psi_j({\bf r})$ is represented by its values at $N$ grid
points, these operations scale with $\mathcal{O}(N)$ and
$\mathcal{O}(N\log N)$, respectively.

The timestep-dependent error $\mathcal{O}(\epsilon^{2})$ introduced by
the decomposition process imposes an upper limit for the time step if
a certain accuracy has to be reached. On the other hand, the whole
procedure needs fewer iterations when the time step is large. To
achieve faster convergence, it would be desirable to use higher order
algorithms at larger time steps.  Unfortunately, Sheng \cite{sheng89}
and Suzuki \cite{nogo} have shown that no factorization of the
evolution operator \eqref{evol} into a product of the operators
$e^{-\epsilon \alpha V}$ and $e^{-\epsilon \beta T}$ beyond second
order can have all positive coefficients $\alpha, \beta$.  This
so-called \emph{forward time step requirement} is, however, essential
for the numerical stability of the algorithm. Fourth order algorithms
with exclusively positive time steps have been derived by Suzuki
\cite{su95,Suzuki96} and Chin \cite{ChinPLA97} by introducing an
additional correction to the potential of the form
$\left[V,\left[T,V\right]\right]$. This correction term has first been
used by Takahashi and Imada \cite{ti84}. We have shown previously
\cite{imstep} that these forward fourth-order algorithms can achieve
similar accuracy at more than an order of magnitude larger step sizes
compared to second-order splitting \eqref{so}.

\subsection {Multi-product expansion}
\label{ssec:any}

Recently, we have made important progress in deriving higher-order
factorization schemes \cite{any,sch_anysolve} by using a linear
combination of second-order propagation steps,
\begin{equation}
 {\cal T}_{2n}\left(\epsilon\right)
\equiv \sum_{k=1}^n c_k {\cal T}_2^k \left(\frac\epsilon{k}
\right)
= {\rm e}^{-\epsilon\left( T+ V + \mathcal{O}(\epsilon^{2n}) \right)}\,.
\label{eq:mcomp}
\end{equation}

The coefficients $c_k$ are given in closed form for any $k$
\cite{chin083}. We have recently implemented a diffusion algorithm
based on such an expansion \cite{any,sch_anysolve}. One $2n$-th order
propagation step $\mathcal{T}_{2n}(\epsilon)$ requires $n(n+1)/2$
second-order propagation steps.  This is more than outweighed by the
fact that much larger timesteps can be used for the higher-order
algorithms due to the smaller time-step error. For instance, we have
shown in Refs.~\cite{any,sch_anysolve} that the 12-th order algorithm,
which requires 21 second-order propagations, yields the same accuracy
as the underlying second-order scheme at a time step about a factor of
1000 larger.  The advantage of high-order propagation methods is
particularly compelling on parallel computer architectures: The
calculation of the action of the evolution operator \eqref{evol} on
the wave functions can be parallelized efficiently by simply
distributing the wave functions $\psi_j({\bf r})$ across different
processors. As higher-order algorithms need much fewer, but more expensive
propagation steps, the relative weight of orthogonalization, which can
be parallelized less efficiently, decreases. Consequently, the
parallel speedup ratio is higher for higher-order algorithms
\cite{any}.

\subsection{Pseudopotentials}
 
Chemical binding properties are almost exclusively determined by the
valence electrons, a fact that is particularly true for metals like
sodium and magnesium where the $s$-type outer electrons are well
separated from a noble gas type ion core. The idea to ignore the
strongly bound core electrons in calculations and to reduce the
nucleus and the core electrons to a ``black box'', an ion core that
interacts with the valence electrons by means of an effective
\emph{pseudopotential,\/} dates back to Fermi \cite{Fermi157}.

Pseudopotentials used in modern electronic structure calculations
generally fall into two categories: One type, {\em empirical\/}
pseudopotentials, uses an analytic model potential with parameters
that are fitted to experimental data. So-called {\em ab-initio\/}
pseudopotentials, on the other hand, are obtained by inverting the
free-atom Schr\"odinger equation for a given reference configuration
\cite{Zunger5449}, and enforce the pseudo-wave functions to coincide
with the true all-electron wave functions outside a given core
radius. The pseudopotentials in this group are usually non-local. The
diffusion algorithm, which was presented for local potentials in
Sec.~\ref{ssec:diffusion}, has also been implemented for non-local
pseudopotentials \cite{nonloc}. The program package \texttt{limerec}
supports both local as well as norm-conserving non-local
pseudopotentials in the form generated by the \textsc{fhi98pp}
\cite{fuchs1999} pseudopotential generator. Most calculations in this
work have been carried out using the local, empirical potentials of
Fiolhais {\em et al.\/} \cite{PerdewPotentials}.  For the purpose of
testing the sensitivity of our results to the pseudopotential used, we
have performed the calculations for Na clusters also with the
potential proposed by K\"ummel {\em et al.\/} \cite{Kueb00}. In one
case, the local potentials predicted a ground state structure
different from what has been reported in the literature. In this case,
we have checked the calculations using non-local Troullier-Martins
pseudopotentials \cite{troullier1993}.

\subsection{Structure optimization}
\label{sec:structureoptimization}

For structure optimization, we have used a simulated annealing
procedure based on Langevin \cite{binggeli92} dynamics of
the ions and a steepest descent method.

The Langevin method is similar to the Metropolis algorithm
\cite{MRRTT53}, but additionally takes into account the influence of
forces on the particles. The ions at positions $\{\mathbf{R}_i \}$ are
moved according to
\begin{equation}
\label{eq:langevin}
{\bf R}_i^{(N+1)} = {\bf R}_i^{(N)} + \delta x\; {\bf g}(i)
- {\beta\over 2}\delta x^2\, \mathbf{F}_i^{(N)}, 
\end{equation}
where ${\bf g}(i)$ is a sequence of {\it Gaussian random numbers},
$\delta x$ is a factor that determines the amplitude of the random
walk of the atoms, $\mathbf{F}_i^{(N)}$ is the force on the $i$-th
atom in simulation step $N$, and $\beta = 1/(k_B T)$, where $T$ is an
artificial temperature. The forces on the ions can be calculated
from a single DFT calculation at fixed ion positions
$\{\mathbf{R}_i\}$ using the Hellman-Feynman theorem
\cite{hellmann1933,feynman1939}.  After each move, the energy
difference
\begin{equation}
\Delta E = E[\{ {\bf R}_i^{(N+1)}\}]-E[ \{{\bf R}_i^{(N)}\}]
\end{equation}
is calculated, where $E[ \{{\bf R}_i^{(N)}\}]$ is the energy of the
cluster at fixed ion positions $\{\mathbf{R}_i\}$ at step $N$, as
calculated by DFT. If $\Delta E < 0$, the move is accepted in any case,
for $\Delta E > 0$, the move is accepted with a probability
$\exp(-\beta \Delta E)$. Compared to the Metropolis algorithm,
Lan\-ge\-vin dynamics allows for higher acceptance rates.

Langevin simulated annealing is guaranteed to converge to the global
energy minimum asymptotically, {\em i.e.,\/} in the limit of
infinitely many annealing steps. For a detailed analysis of the
convergence properties of the meth\-od, including the finite-time
behavior, see Ref.~\cite{MITRA:1986vj}. In practical calculations,
cooling the system faster or terminating the annealing process earlier
leads to an increased probablility of freezing it in a higher-lying,
local energy minimum. One way to verify the robustnes of the procedure
is to repeat it from different starting configurations.  To find the
ground-state configurations and isomers reported in this work, we have
thus proceeded as follows:
\begin{enumerate}
\item We have generated a number of random starting configurations of
  ion core positions for each cluster, only ensuring that the
  distances between the ion cores are large enough to avoid
  divergences.
\item For each starting configuration, we have performed a
  simulated annealing using Langevin molecular dynamics: After a
  suitable thermalisation period, the temperature in
  Eq.~\eqref{eq:langevin} was slow\-ly decreased to ``freeze'' the
  cluster in a low-lying minimum of the energy surface.
\item As soon as the annealing procedure had settled close to a local
  minimum at low temperature, we switched to a steepest descent
  method to accurately determine the position of the minimum. This was
  implemented by using the Verlet molecular dynamics method for the
  ion cores and setting, after each step, all ion velocities to
  zero. This changed the ion positions by typically about $0.1\,a_0$.
\item If only ground-state configurations are of interest, the
  annealing process can in principle be stopped as soon as it becomes
  clear that it converges towards a higher energy than a previously
  found configuration. To find isomers, we have nevertheless
  completed the full annealing procedure for all Na$_n$ and Na$_n^+$
  clusters shown here, as well as for Mg$_n$ and Mg$_n^+$ up to $n=15$.
\item All found configurations were checked for congruency, {\em
    i.e.,\/} for being identical up to rotations and/or translations.
  It turned out that all structures with an energy difference greater
  than $10^{-4}\,$Ry per electron were isomers. In four cases we have
  also found incongruent configurations with an energy difference of
  \emph{less} than $10^{-4}\,$Ry per electron. Two of these will be
  discussed below.
\end{enumerate}

To estimate the discretization error, we have also calculated, for
some sample clusters, the total energy while rigidly moving and
rotating the ions in different directions in the simulation box. For
all clusters, the discretization error is around $10^{-5}$ Ry, which
is certainly much smaller than the uncertainty implicit to the local
density approximation.

In order to find a locally stable configuration, we have typically
carried out about 1000 Langevin moves, slowly decreasing the
artificial temperature $T = 1/ (k_B \beta )$.  These were followed by
several hundred steepest descent moves. Therefore, each locally stable
configuration shown here involved between 1000 and 2000 LDA
calculations for the cluster under consideration. The diffusion
algorithm described above is particularly efficient for such a
strategy, because it iteratively improves a given set of initial wave
functions, and thus profits from good starting values. Since the ions
are moved only very little during one annealing step, the wave
functions do not change much, and the diffusion algorithm for
subsequent steps converges within just a few iterations.

\section{Results}
\label{sec:results}

\subsection{Na and Na$\mathbf{^+}$  clusters}
\label{ssec:Na}

Na$_n$ and Na$_n^+$ clusters are the most frequently studied type of
metallic clusters, for a recent review see Ref. \cite{IKAK09}. We have
examined these systems basically to highlight the differences between
Na$_n$ and Mg$_n$ clusters.  We have performed (S)DFT calculations for
small Na$_n$ and Na$_n^+$ clusters up to $n = 10$. The calculations
were carried out on a regular, Cartesian real-space grid with 64$^3$
mesh points and a resolution of 0.5$\,\mathrm{a_0}$, where
$\mathrm{a_0}$ is the Bohr radius. We have determined that this is the
coarsest discretization that is permitted to get reliable results. Our
discretization corresponds to an energy cutoff of 537$\,$eV in plane-wave
calculations.

All calculations in this section have been carried out using the
empirical pseudopotentials of both Fiolhais {\em et al.\/}
\cite{PerdewPotentials} and K\"ummel {\em et al.\/} \cite{Kueb00}, for
the purpose of testing the sensitivity of the results to the
pseudopotential used. In one case, we have additionally used
norm-conserving non-local pseudopotentials of the Troullier-Martins
type \cite{troullier1993}, see Fig.~\ref{fig:na-compare_04}.

\subsubsection{Energetics}

The binding energy for neutral and singly charged clusters is defined by
\begin{equation}
 \label{eq:bindenergy}
 \begin{split}
  E_{\mathrm{b},n}^{\phantom{+}} &\equiv nE_1 - E_n\,,\\
  E_{\mathrm{b},n}^+ &\equiv (n-1)E_1 + E_1^+ - E_n^+\,,
 \end{split}
\end{equation}
where $E_1$ and $E_n$ are the ground state energies of a single Na
atom and a Na$_n$ cluster, respectively, and $E_1^+$ and $E_n^+$ are
their counterparts for the charged clusters.

\begin{figure}
  \centering
  \includegraphics[width=\linewidth, keepaspectratio]{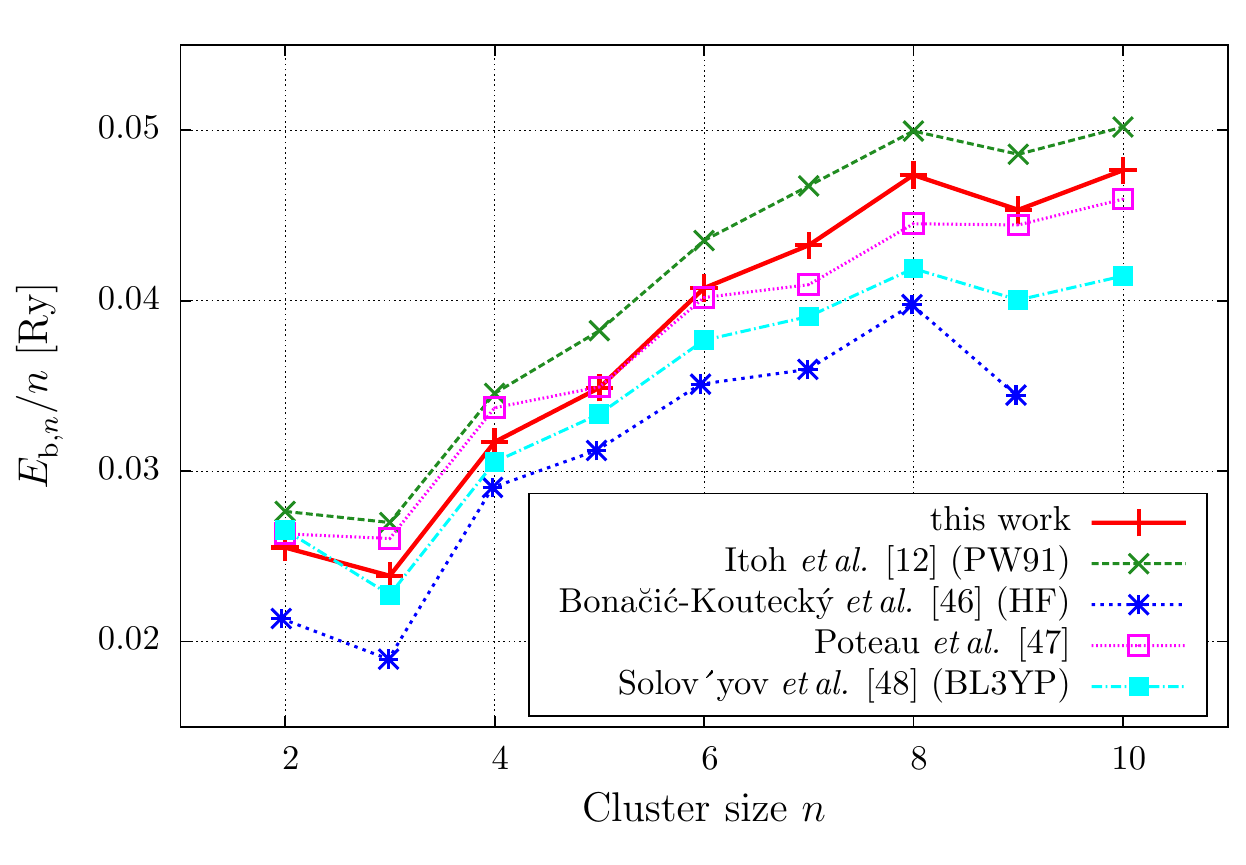}
  \caption{(color online) The binding energy per atom of neutral Na$_n$
    clusters compared with earlier work as cited in the legend. The
    results of this work have been obtained using the potentials of
    Fiolhais {\em et al.\/} \cite{PerdewPotentials} and those by
    K\"ummel~{\it et al.\/}~\cite{Kueb00}. }
  \label{fig:na-bind-energy_neut}
\end{figure}

\begin{figure}
  \centering
  \includegraphics[width=\linewidth, keepaspectratio]{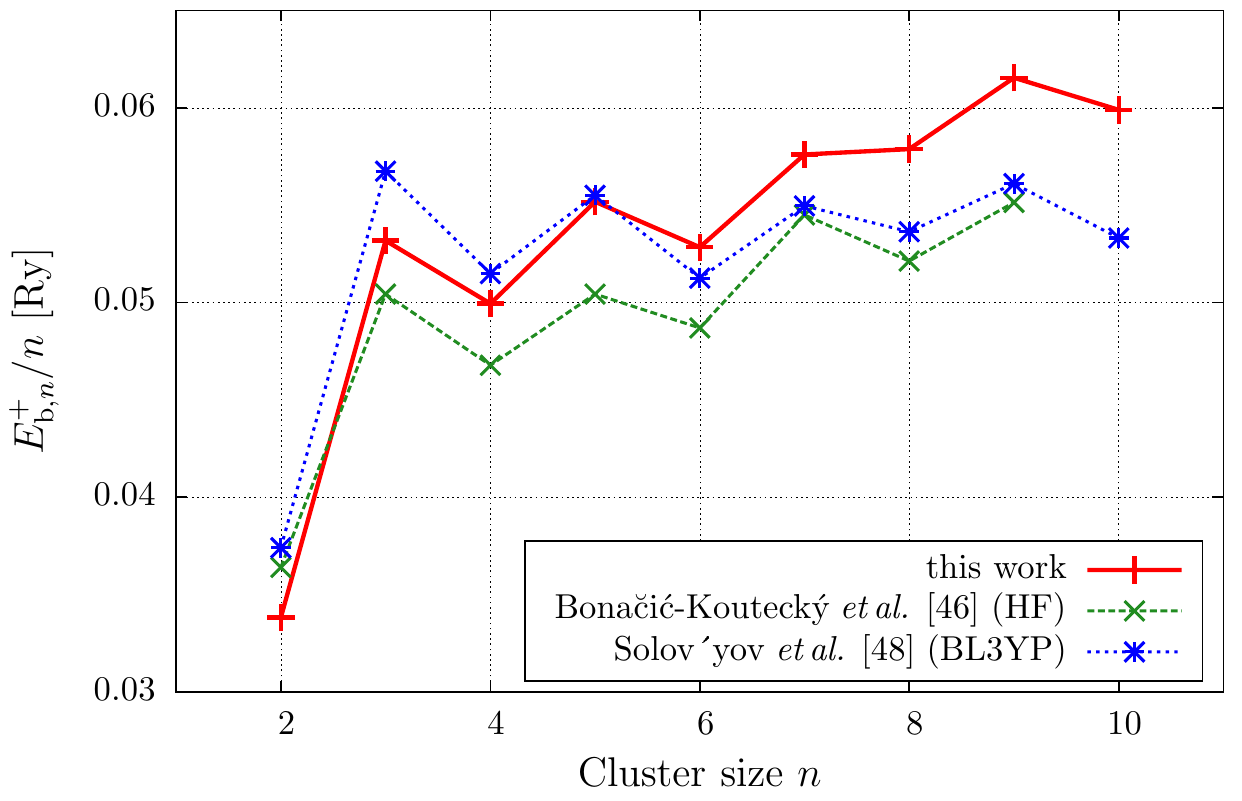}
  \caption{(color online) Binding energy per atom of singly ionized
    Na$_n^+$ clusters compared with earlier work as cited in the
    legend. The results of this work have been obtained using the
    potentials of Fiolhais {\em et al.\/} \cite{PerdewPotentials} and
    those by K\"ummel~{\it et al.\/}~\cite{Kueb00}. }
  \label{fig:na-bind-energy_plus}
\end{figure}

Figs.~\ref{fig:na-bind-energy_neut} and \ref{fig:na-bind-energy_plus}
show the binding energy per atom for Na$_n$ and Na$_n^+$ clusters in
comparison with previous work
\cite{IKAK09,bonacic4861,Poteau1878,Solovyov053203}, where we find
generally good agreement.  For neutral clusters we have found that the
pseudopotential proposed by Fiolhais {\em et al.\/}
\cite{PerdewPotentials} and that of K\"ummel~{\it et
  al.\/}~\cite{Kueb00} give almost identical results. For charged
clusters, the difference in the binding energy due to different
pseudopotentials is comparable to the discrepancy between different
calculations.

Consistent with earlier work, we observe a particularly high stability
for Na$_2$, Na$_8$, Na$_3^+$ and Na$_9^+$, where the binding energy
per atom is higher than the one of the cluster with one additional
atom. This is known to be due to electronic shell closure and
predicted by ellipsoidal jellium models~\cite{deHeerRMP}. We also
observe an additional peak for Na$_5^+$ which can not be explained by
shell closures. This peak is consistent with earlier theoretical
calculations \cite{Tevekeliyska06}; it has also been observed in
Na$_n^+$ fragmentation experiments \cite{Baumert92}. It is the onset
of an even-odd alternation which has been attributed to electron
pairing effects \cite{Ekardt88}.


\subsubsection{Geometric structure}

\begin{figure}
  \centering
  \includegraphics[width=\linewidth,keepaspectratio]{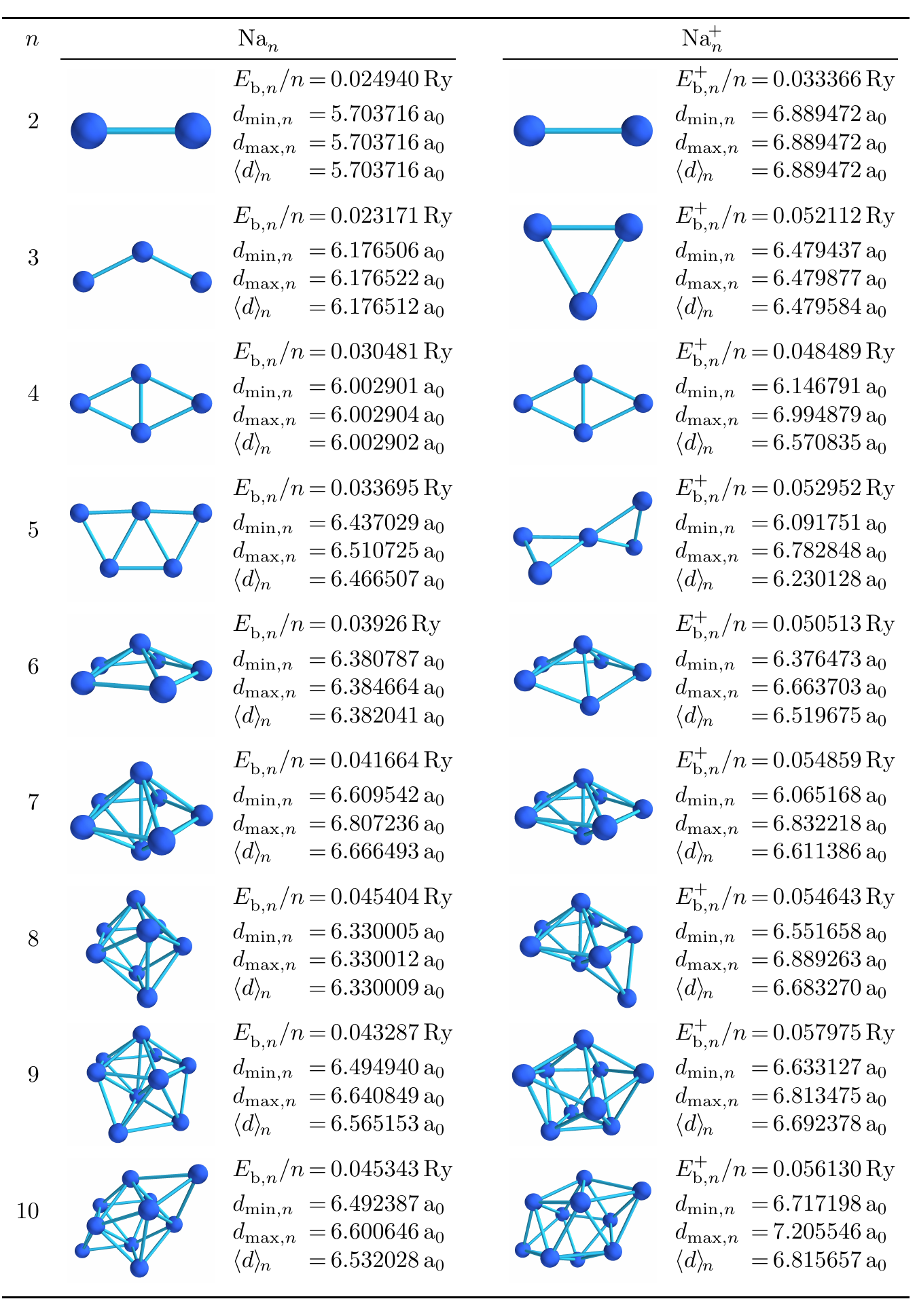}
  \caption{(color online) Ground state configurations of neutral
    Na$_n$ and singly charged Na$^+_n$ clusters up to $n=10$. The results
    reported here were obtained using the potential of Fiolhais {\em
      et al.\/} \cite{PerdewPotentials}. Note that we show, in the
    case Na$_4$, the rhomboid isomer. Local pseudopotentials predict
    an additional distorted configuration slightly lower in energy,
    see Fig.~\ref{fig:na-compare_04}. $E_{b,n}/n$ is the binding
    energy per atom, $\langle d\rangle_n$ the average nearest
    neighbor distance, and $d_{\mathrm{min},n}$ and
    $d_{\mathrm{max},n}$ are the smallest and largest nearest neighbor
    distances, respectively.}
  \label{fig:na-clusters}
\end{figure}

An overview of the annealed ground state structures for Na$_n$ and
Na$_n^+$ is given in Fig.~\ref{fig:na-clusters}.

\begin{figure}
  \centering
  \includegraphics[width=0.45\linewidth,keepaspectratio]{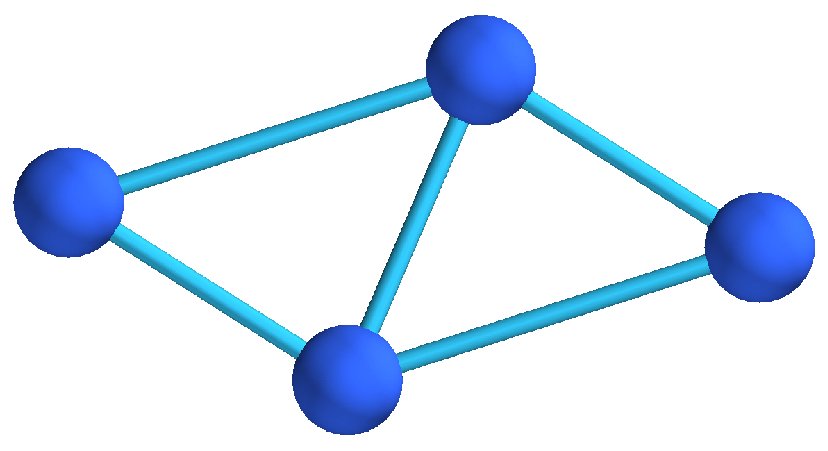}
  \hspace{0.05\linewidth}
  \includegraphics[width=0.45\linewidth,keepaspectratio]{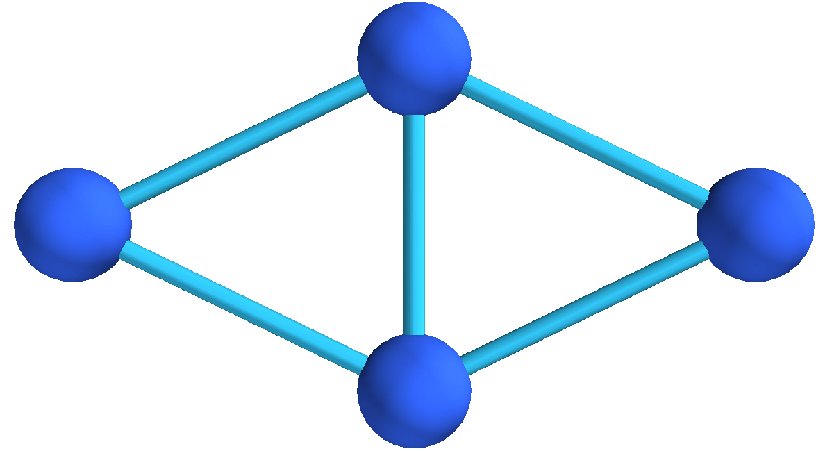}
  \caption{(color online) Ground state configurations of Na$_4$. Left pane:
    ``distorted'' rhomboidal configuration which is found as the ground
    state with both local pseudopotentials. Right pane: Symmetric
    configuration, found as slightly higher-lying isomer with local potentials
    (energy difference: $+3.96\cdot10^{-4}\,\mathrm{Ry}$). For
    non-local Troullier-Martins pseudopotentials, the distorted
    configuration is not stable, and the symmetric configuration shown
    right is the ground state structure.}
  \label{fig:na-compare_04}
\end{figure}

  The agreement with earlier work is consistent with what is expected
  from the energetics. The only noticeable deviation from literature
  has been observed for Na$_4$, where the energy minimization with
  both local pseudopotentials converges towards a distorted rhomboidal
  structure, see Fig.~\ref{fig:na-compare_04}. The symmetric
  rhomboidal structure, which is usually reported as the ground state
  in the literature (see, for instance, Ref.~\cite{Solovyov053203})
  appears as an isomer only $3.96\cdot10^{-4}\,\mathrm{Ry}$ higher in
  energy. When using non-local pseudopotentials of the
  Troullier-Martins type \cite{troullier1993}, the distorted structure
  ceases to be a local minimum, and the symmetric configuration is
  found for the ground state. The effect thus seems to be connected to
  the locality of the potential.

\subsubsection{Kohn-Sham energies}

\begin{figure}
  \centering
  \includegraphics[width=\linewidth,keepaspectratio]{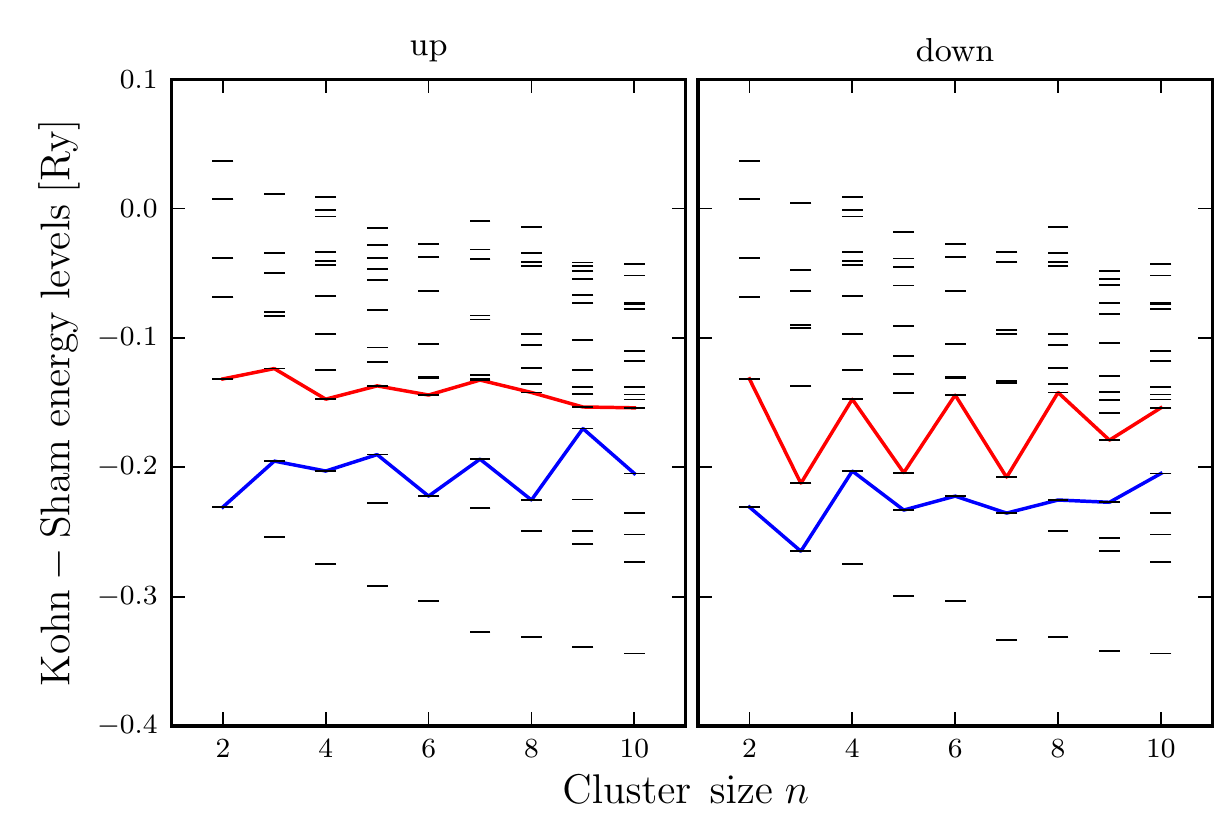}
  \caption{(color online) Kohn-Sham energy levels of the neutral
    Na$_{n}$ clusters as a function of cluster size. Left pane: Energy
    levels for spin-up electrons. Right pane: Energy levels for
    spin-down electrons.  The HOMO levels are marked by the lower
    (blue) solid line, the LUMO levels by the upper (red) solid
    line. The even electron numbers have been calculated with LDA, the
    odd ones with LSDA, hence the energy levels of spin-up and
    spin-down electrons of the configurations with even electron
    numbers are the same.}
  \label{fig:na-KS_levels_neut}
\end{figure}

\begin{figure}
  \centering
  \includegraphics[width=\linewidth,keepaspectratio]{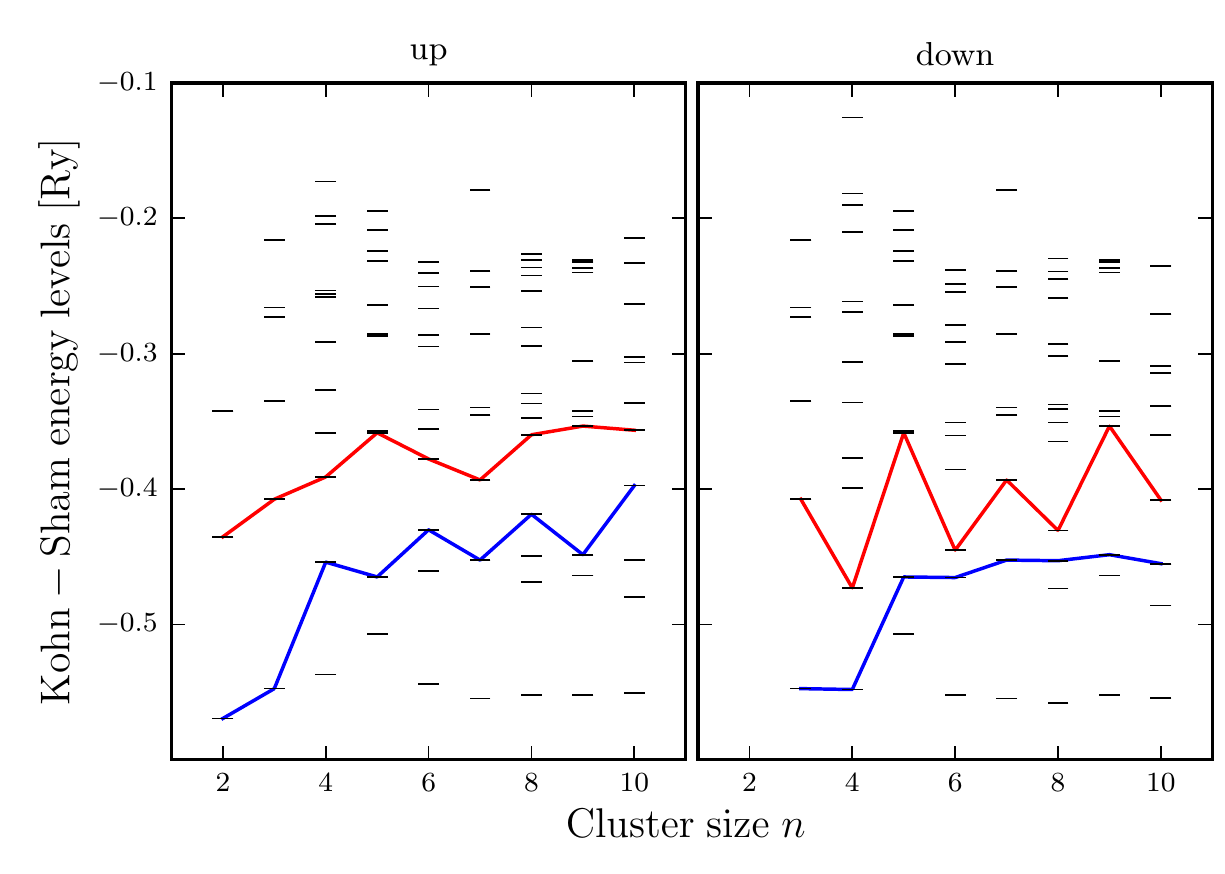}
  \caption{(color online) Kohn-Sham energy levels of the singly
    ionized Na$_{n}^{+}$ clusters as a function of cluster size. See
    Fig.~\ref{fig:na-KS_levels_neut} for further explanations.}
  \label{fig:na-KS_levels_plus}
\end{figure}

The Kohn-Sham energies of the ground state structures for Na$_n$ and
Na$_n^+$ are depicted in Figs.~\ref{fig:na-KS_levels_neut} and
\ref{fig:na-KS_levels_plus}. For odd electron numbers we have used the
LSDA, which implies that electrons with different spins can have
different wave functions and energy eigenstates. The lowest states in
these figures correspond to the 1$s$-states. The 1$p$-states are
separated by a pronounced gap and can be clearly identified, as well
as the onset of the 1$d$-states.

The energy levels show a good overall agreement with the
single-particle levels predicted by the Clemenger-Niels\-son model
\cite{cle1359}, see Fig.~5(a) in Ref.~\cite{deHeerRMP}. This indicates
that ellipsoidal jellium models give a good qualitative description of
Na clusters.  One noteable difference is Na$_8$: While clusters with
closed shells usually are spherically symmetric and thus show
degenerate $p$-states in jellium models, this degeneracy has been
lifted in our calculation. This is a result of the symmetry breaking
due to the ion cores, and has been observed previously, see, e.g.,
Fig.~5(d) in Ref.~\cite{deHeerRMP}.

The Clemenger-Nilsson model predicts shell closures for Na$_2$,
Na$_8$, Na$_3^+$ and Na$_9^+$
\cite{deHeerRMP,cle1359}. Correspondingly, the neighboring clusters
show a significantly decreased gap in at least one spin channel in
Figs. \ref{fig:na-KS_levels_neut} and \ref{fig:na-KS_levels_plus}. In
addition to these shell closures, we also observe the even-odd
alternation attributed to the electron pairing effects mentioned above
\cite{Ekardt88}. Due to this effect, also Na$_6$ and Na$_5^+$ show a
particularly large gap.


\subsection{Mg and Mg$\mathbf{^+}$ clusters}
\label{ssec:Mg}

The focus of our work is the calculation of structures and energetics
of Mg$_n$ and Mg$_n^+$ clusters. We have again used the procedure
outlined in Sec.~\ref{sec:structureoptimization} to find ground
states and low-lying isomers of Mg$_n$ and Mg$_n^+$ clusters up to
$n=30$.  The ion cores of the Mg atoms have been described by the
local potentials of Ref.~\cite{PerdewPotentials}. 

Due to the shorter range of the Mg pseudopotentials compared to those
for Na, all calculations have been performed on a regular, Cartesian
real-space grid with 96$^3$ mesh points and a resolution of
0.375$\,\mathrm{a_0}$. This corresponds to an energy cutoff of
954.5$\,$eV in a plane-wave calculation.

A comment is in order concerning the expected accuracy of the applied
theory. This can be assessed by comparing predictions for small
systems with results obtained by more elaborate
methods. Refs.~\cite{Jellinek02} and \cite{Koehn01}, for instance,
compare the bond lengths of Mg$_3$ and Mg$_4$ obtained from coupled
cluster (CC) and DFT calculations. The CC calculations included single
and double excitations as well as perturbative triples [CCSD(T)], the
DFT calculations were all-electron calculations using the BP86
correlation functional. The results of both methods agree with ours
within a few percent. This is certainly better than expected because
the pseudopotentials used in our work have been designed to reproduce
bulk properties.

\subsubsection{Energetics}
\label{sec:energy}

To safely bracket the regime where a non-metal to metal (NMM)
transition has been proposed \cite{DDB01,diederichPRA05}, we have
calculated structures up to $n=30$.  The binding energy for neutral
and singly charged clusters is defined in Eq.~\eqref{eq:bindenergy},
where $E_1$ and $E_n$ are the ground state energies of a single Mg
atom and a Mg$_n$ cluster, and $E_1^{+}$ and $E_n^{+}$ their
counterparts for the charged clusters, respectively. By repeating the
annealing procedure with different initial configurations, we have
also calculated isomers for ion numbers $n\le 15$; for larger clusters
the number of isomers grows rapidly and identifying all of them
becomes extremely time consuming.

\begin{figure}
  \centering
  \includegraphics[width=\linewidth, keepaspectratio]{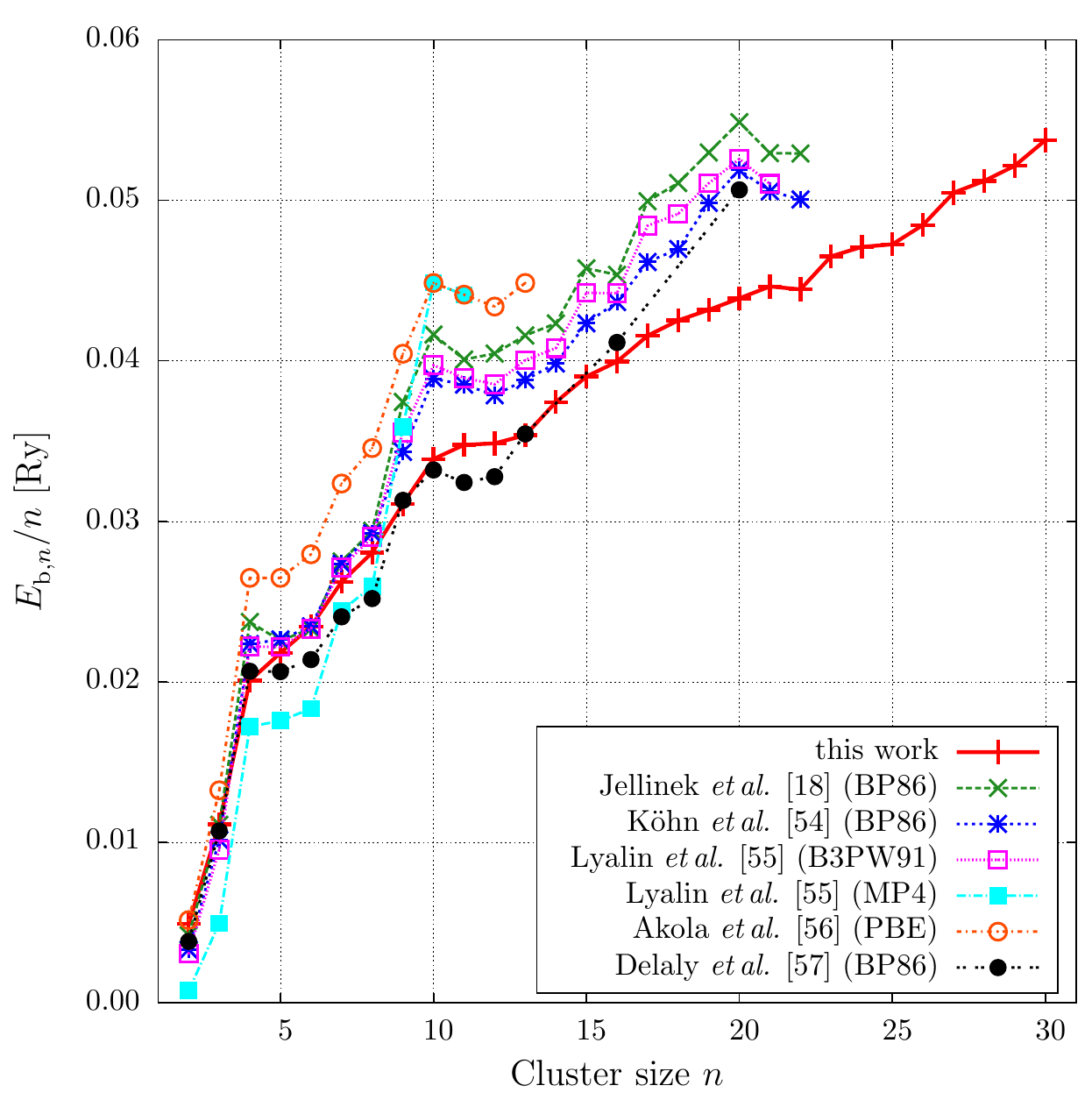}
  \caption{(color online) Binding energy per atom, $E_{b,n}/n$, of
    Mg$_n$ clusters in comparison with earlier calculations as cited
    in the legend.}
  \label{fig:mg-bind-energy_neut}
\end{figure}

\begin{figure}
  \centering
  \includegraphics[width=\linewidth, keepaspectratio]{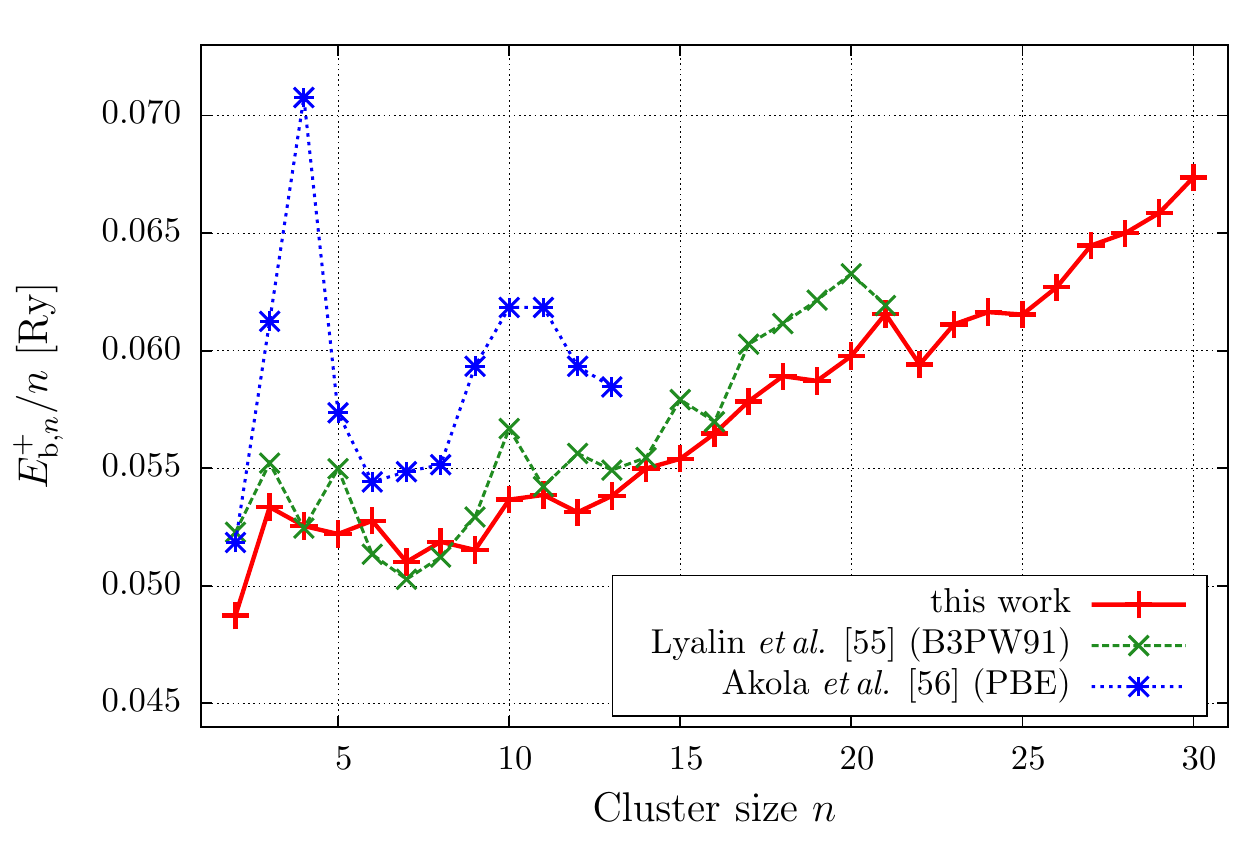}
  \caption{(color online) Binding energy per atom,
    $E_{b,n}^{+}/n$, of Mg$_n^+$ clusters in comparison with earlier
    calculations as cited in the legend.}
  \label{fig:mg-bind-energy_plus}
\end{figure}

Figures \ref{fig:mg-bind-energy_neut} and
\ref{fig:mg-bind-energy_plus} show our results for the binding energy
per atom of the Mg$_n$ and Mg$_n^+$ clusters, compared to a number of
previous calculations
\cite{Jellinek02,Koehn01,LSSG03,Akola01,Del3838}. The deviations
between the calculations are due to the use of different energy
functionals and pseudopotentials, and give an idea of the expected
accuracy of the underlying physical model. The most apparent
discrepancy between our results and earlier work is the greater
smoothness of the binding energy as a function of the number of
atoms. This is likely due to our elaborate annealing procedure in
combination with high spatial resolution.

\begin{figure}
  \centering
  \includegraphics[width=\linewidth, keepaspectratio]{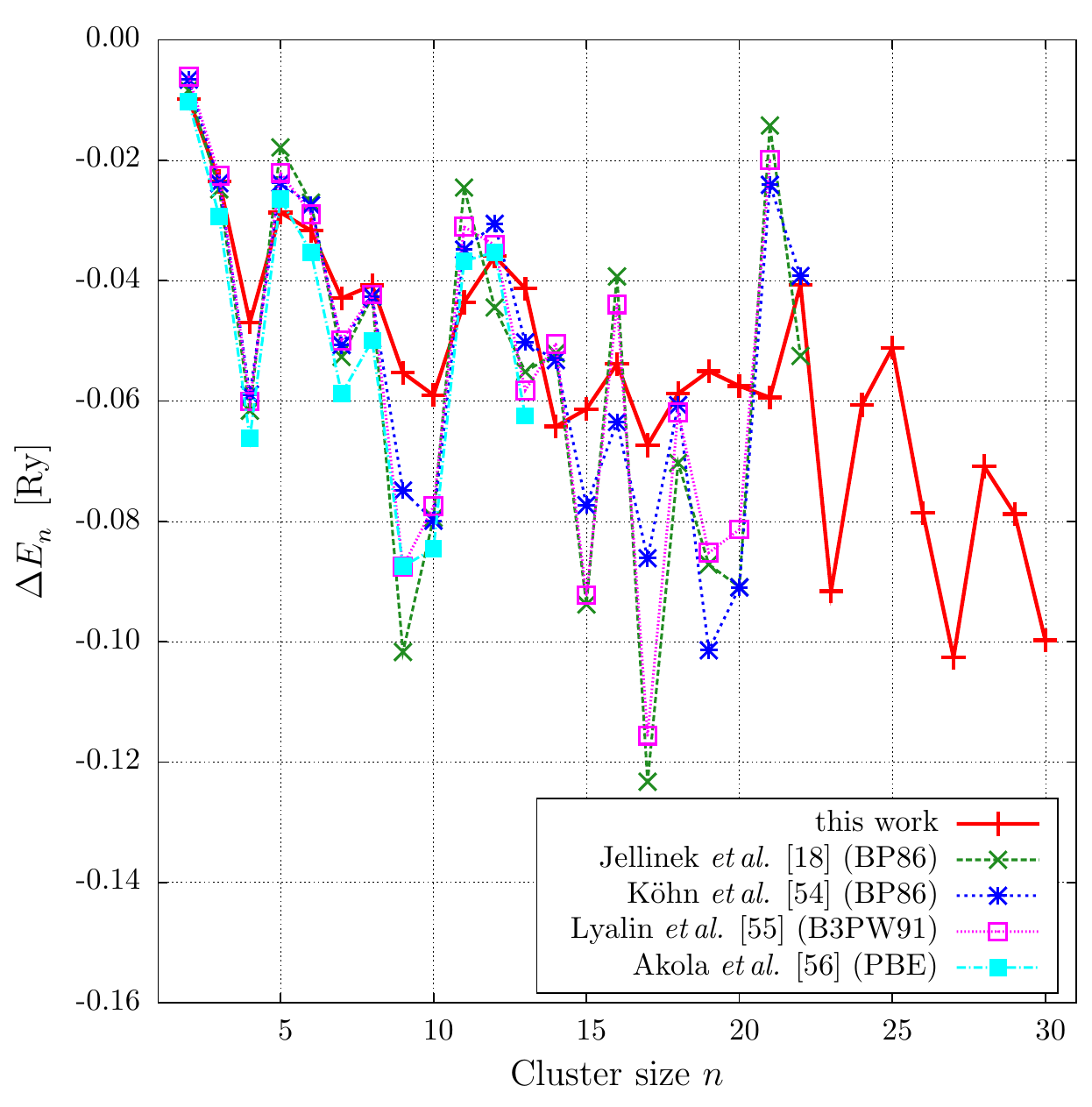}
  \caption{(color online) Fragmentation energies of Mg$_n$
    clusters in the energetically preferred channel, $\Delta E_n$, for
    neutral Mg$_n$ clusters up to $n=30$, see Eq.~\eqref{Eq:deltaE_neutral}.
    Results of some earlier works are shown as cited in the legend.}
  \label{fig:mg-deltae_neut}
\end{figure}

\begin{figure}
  \centering
  \includegraphics[width=\linewidth, keepaspectratio]{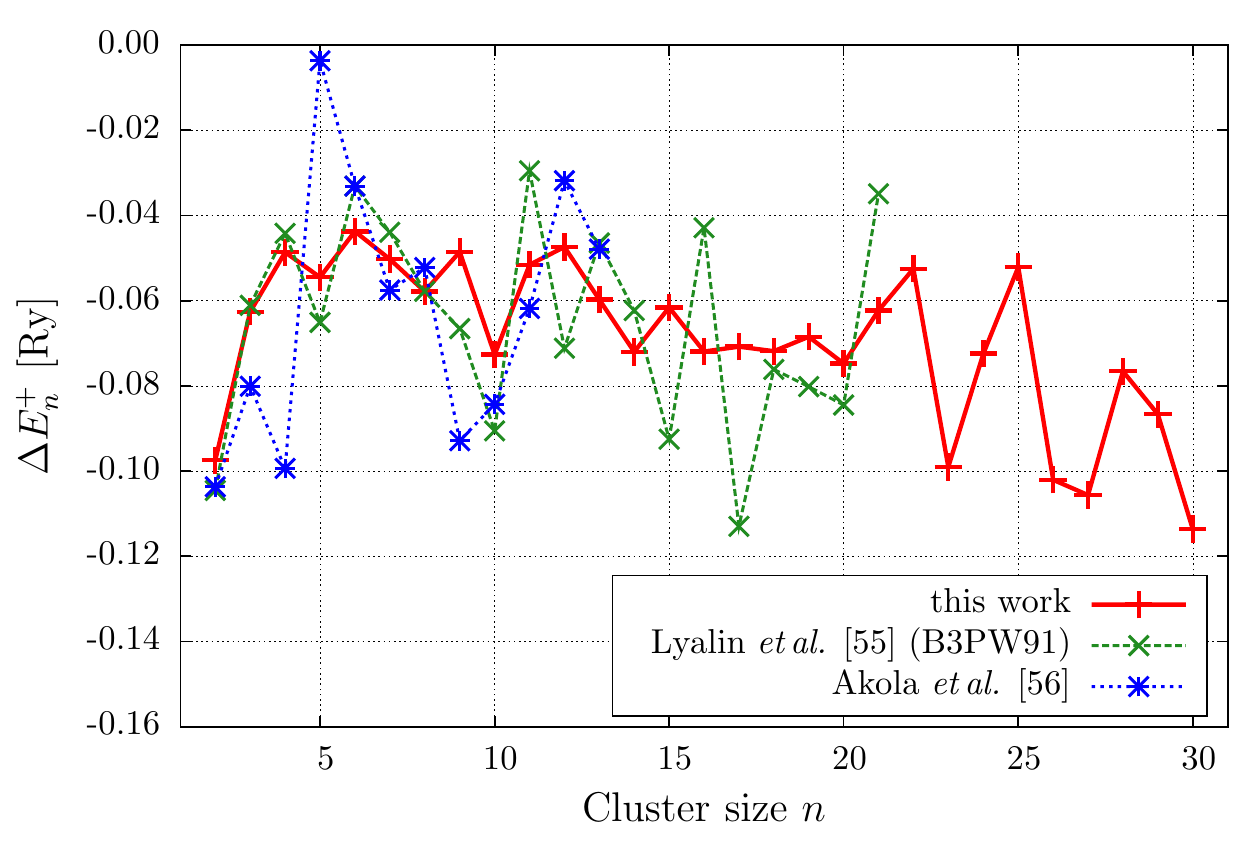}
  \caption{(color online) Fragmentation energies of Mg$_n^+$
    clusters in the energetically preferred channel
    for singly ionized Mg$_n^+$ clusters, $\Delta E_n^+$, up to $n=30$, see
    Eq.~\eqref{Eq:deltaE_neutral}. Results of some earlier works are shown as
    cited in the legend.}
  \label{fig:mg-deltae_plus}
\end{figure}

For a quantitative assessment of cluster stability, we turn our
attention to the possible fragmentation channels. A Mg$_n$ cluster can
dissociate into two smaller clusters of sizes $n-m$ and $m$. The
fragmentation energies are defined as
\begin{equation}
 \begin{split}
  \Delta E_n^{\phantom{+}} &\equiv \max_{1 \leq m < n}
  \big[ E_{\mathrm{b},n-m}^{\phantom{+}} + E_{\mathrm{b},m}^{\phantom{+}} -
  E_{\mathrm{b},n}^{\phantom{+}} \big]\,,\\
  \Delta E_n^{+} &\equiv \max_{1 \leq m < n} \big[ E_{\mathrm{b},n-m}^{+} +
  E_{\mathrm{b},m} - E_{\mathrm{b},n}^{+}, \\
   & \phantom{:= \max_{1 \leq m < n} \big[}\ E_{\mathrm{b},n-m}^{\phantom{+}}
   + E_{\mathrm{b},m}^{+} - E_{\mathrm{b},n}^{+} \big]\,.
 \end{split}
 \label{Eq:deltaE_neutral}
\end{equation}
If this quantity is negative, the cluster is stable against
spontaneous fragmentation.  Nevertheless, for each cluster, we denote
the channel with the highest fragmentation energy as the preferred
fragmentation channel, even if the cluster is stable.
Figs. \ref{fig:mg-deltae_neut} and \ref{fig:mg-deltae_plus} show the
dependence of the fragmentation energy on cluster size.  We found that
all Mg and Mg$^+$ clusters are stable.  The preferred fragmentation
channel for neutral clusters is always
\begin{equation}
 \mathrm{Mg}_n^{\phantom{+}} \longrightarrow
\mathrm{Mg}_{n-1}^{\phantom{+}} + \mathrm{Mg}_1^{\phantom{+}}\,.
\end{equation}
For singly ionized clusters it is, correspondingly,
\begin{equation}
 \mathrm{Mg}_n^+ \longrightarrow \mathrm{Mg}_{n-1}^+ + \mathrm{Mg}_1^{\phantom{+}}
\end{equation}
for all cluster sizes. This is due to the monotonic increase of the
binding energy with the number of atoms, in contrast to Na clusters,
which often decay as $\mathrm{Na}_n^{\phantom{+}} \rightarrow
\mathrm{Na}_{n-2}^{\phantom{+}} + \mathrm{Na}_2^{\phantom{+}}$.


\begin{figure}
  \centering
  \includegraphics[width=\linewidth, keepaspectratio]{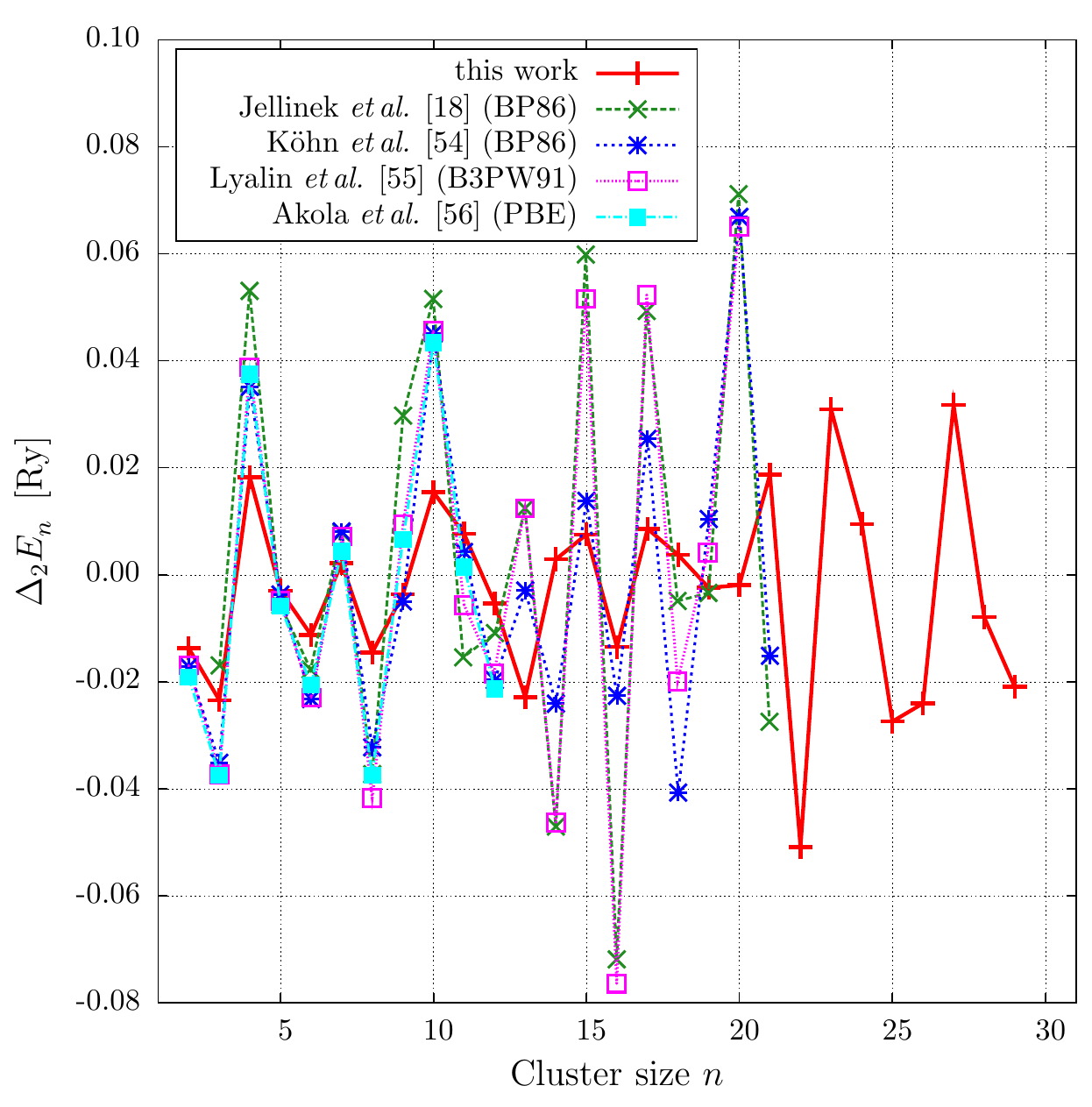}
  \caption{(color online) Second energy differences of neutral
    Mg$_n$ clusters, $\Delta_2 E_n$, up to $n=29$, see Eq.~\eqref{Eq:delta2E}.
    A number of different results reported in the literature is shown.}
  \label{fig:mg-delta2e_neut}
\end{figure}

\begin{figure}
  \centering
  \includegraphics[width=\linewidth, keepaspectratio]{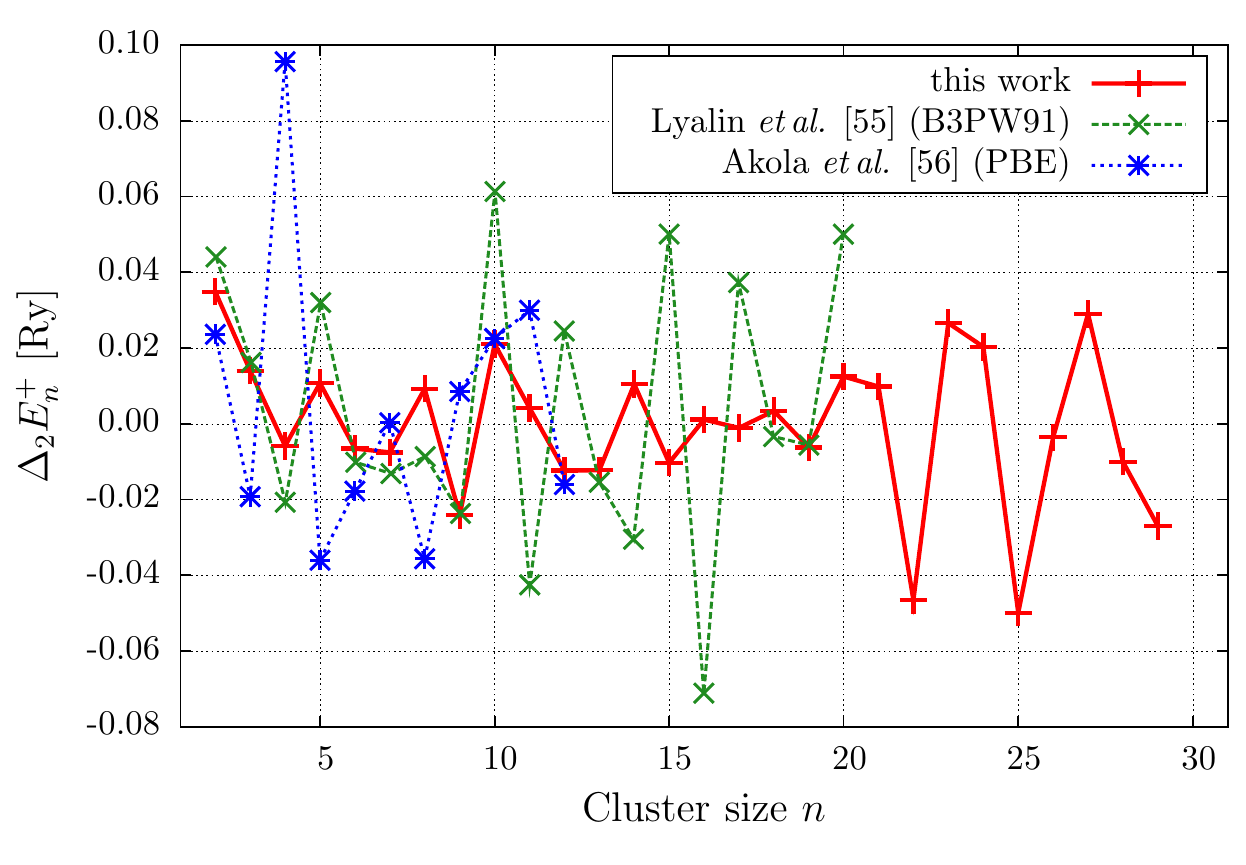}
  \caption{(color online) Second energy differences of singly ionized
    Mg$_n^+$ clusters, $\Delta_2 E_n^+$, up to $n=29$, see Eq.~\eqref{Eq:delta2E}.
    A number of different results reported in the literature is shown.}
  \label{fig:mg-delta2e_plus}
\end{figure}

Another criterion for cluster stability \cite{deHeerRMP}
involves the second energy difference,
\begin{equation}
 \Delta_2 E_n^{(+)} \equiv E_{n-1}^{(+)} -2E_n^{(+)} + E_{n+1}^{(+)}\,,
 \label{Eq:delta2E}
\end{equation}
where the superscript $(+)$ indicates that the formula applies to both
neutral and charged clusters. Eq.\ \eqref{Eq:delta2E} can be understood
as a discretized second derivative of the electronic energy with
respect to the ion number.  If $\Delta_2 E_n^{(+)}>0$, then
Mg$_n^{(+)}$ is stable with respect to disproportionation of two
Mg$_{n}^{(+)}$ clusters into Mg$_{n-1}^{(+)}$ and Mg$_{n+1}^{(+)}$.  A
higher value indicates a more stable cluster. The evolution of these
quantities with cluster size among some representative results found
in the literature are shown in Figs. \ref{fig:mg-delta2e_neut} and
\ref{fig:mg-delta2e_plus}, respectively.


We can use both the fragmentation energies {\em and\/} the second
energy differences to identify extraordinarily stable clusters. To
find those, we look for clusters that show a distinct minimum in
$\Delta E_n^{(+)}$ \emph{and} a distinct maximum in $\Delta_2
E_n^{(+)}$ in the region where $\Delta_2 E_n^{(+)}>0$. This procedure
suggests that neutral Mg$_n$ clusters with 4, 7, 10, 17, 21, 23 and 27
atoms are particularly stable. For the ionized Mg$_n^+$ clusters, the
most stable sizes are 5, 8, 10, 14, 20, 23 and 27,
respectively. Diederich {\em et al.\/} \cite{diederichPRA05} have
measured the abundance of Mg$_n^+$ clusters grown in ultracold liquid
helium nanodroplets using mass spectroscopy.  They report, in the area
of cluster sizes covered here, distinct maxima in the abundance
distribution at clusters sizes 5, 10 and 20, as well as a local
maximum at 8. All of these ``magic numbers'' are consistent with the
stable cluster sizes predicted by the above procedure.

The clusters Mg$_{14}^+$,
Mg$_{23}^+$ and Mg$_{27}^+$, which are also predicted to be particularly
stable by our analysis, do not show up as abundancy maxima in the
experiment. Two of these clusters, Mg$_{14}^+$ and Mg$_{23}^+$, seem
to have a somewhat peculiar geometrical structure, see
Sec.~\ref{sec:mgstructure}.

\begin{figure}
  \centering
  \includegraphics[width=\linewidth, keepaspectratio]{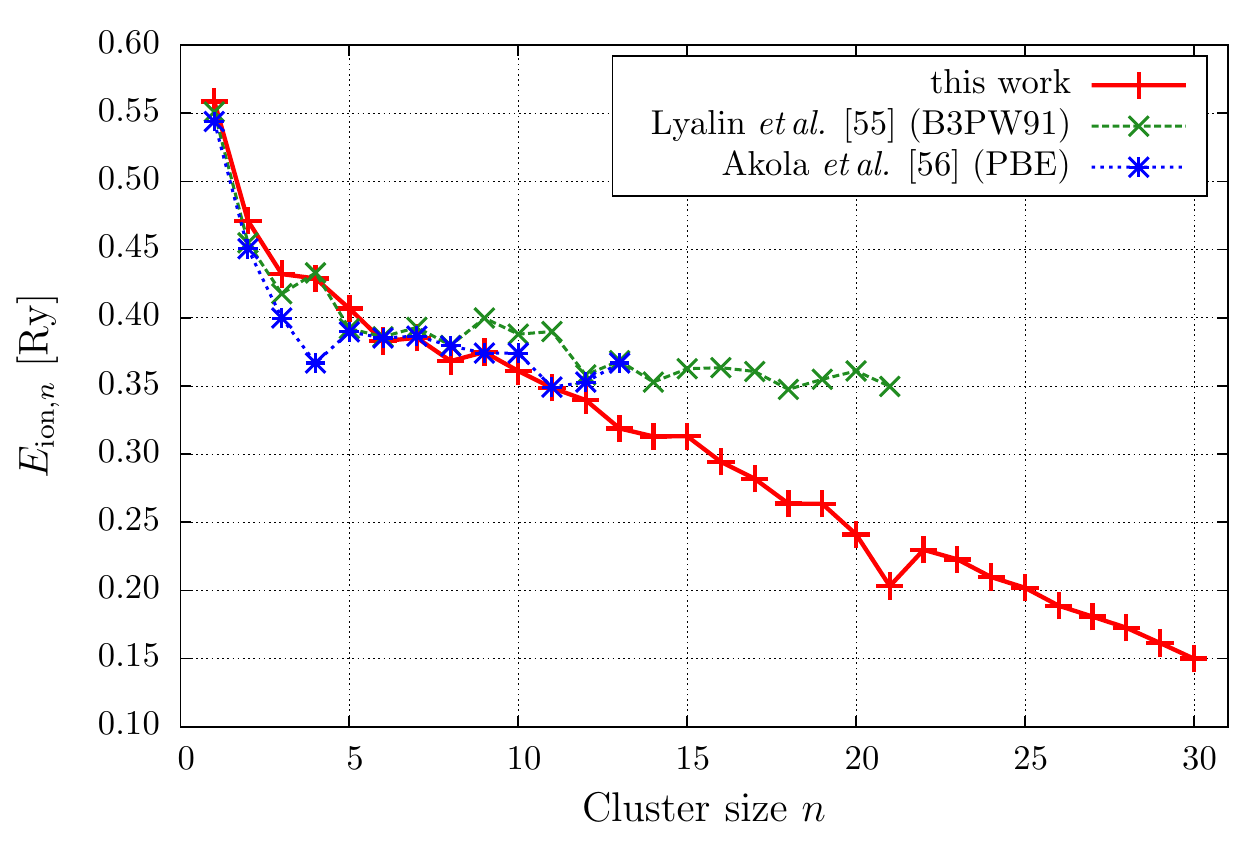}
  \caption{(color online) Adiabatic ionization energies, $E_{\mathrm{ion},n}$, of Mg
   clusters up to $n=30$ in comparison with earlier calculations as
   cited in the legend.}
  \label{fig:mg-ionization}
\end{figure}

The evolution of the adiabatic ionization energies,
\begin{equation}
 E_{\mathrm{ion},n} \equiv E_n^+ - E_n^{\phantom{+}}\,,
 \label{Eq:ionization}
\end{equation}
with cluster size is shown in Fig.~\ref{fig:mg-ionization}. The
almost monotonically decreasing values are in good agreement with
previous calculations \cite{LSSG03,Akola01}.

\subsubsection{Geometric structure}
\label{sec:mgstructure}

\begin{figure}
  \centering
  \includegraphics[width=\linewidth,keepaspectratio]{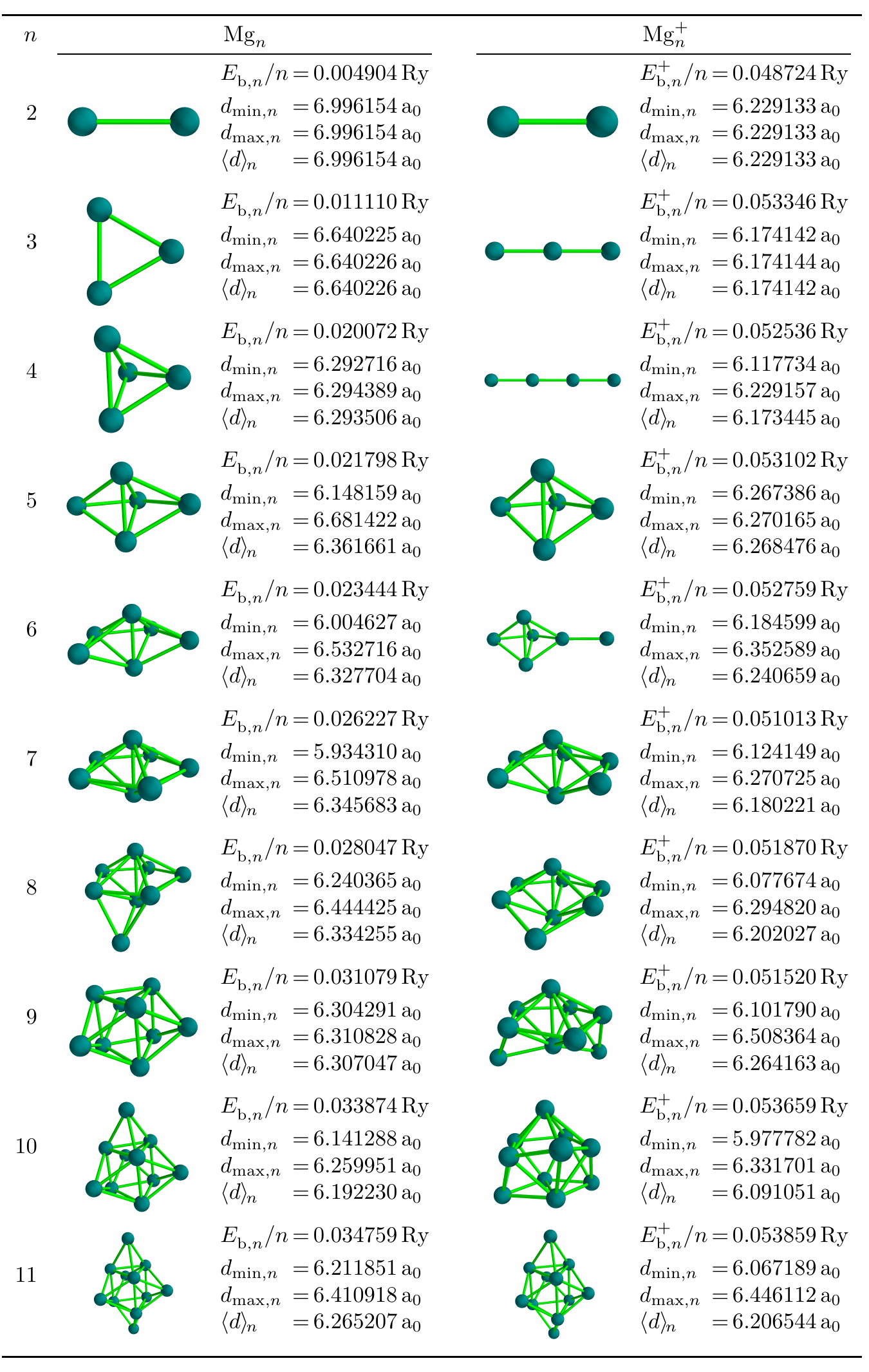}
  \caption{(color online) Ground state configurations of neutral
    Mg$_n$ and singly charged Mg$^+_n$ clusters up to $n=11$. The ground state
    structures of all Mg$_n$ clusters with $n \le 30$ can be found in
    Ref.~\cite{liebthes}; isomers will be discussed in
    Sec.~\ref{sec:isomers}\,. $E_{b,n}/n$ is the binding energy per
    atom, $\langle d\rangle_n$ the average nearest neighbor
    distance, and $d_{\mathrm{min},n}$ and $d_{\mathrm{max},n}$ are the smallest and largest nearest neighbor distances, respectively.
  \label{fig:mg-clusters}}
\end{figure}

Fig.~\ref{fig:mg-clusters} shows the ground state
geometries of the neutral and charged Mg$_n$ clusters up to
$n=11$.
Displaying the individual ion positions for larger clusters does not
provide much useful information, these positions can be found in
Ref. \cite{liebthes}.

To examine the structure of larger clusters, we look at the pair
distribution function of the ions.  The classical two-body density of
the ion cores,
\begin{equation}
 \rho_2 ({\bf r},{\bf r}') \equiv
 \sum_{i \neq j}\delta ({\bf r}-{\bf R}_i)
\delta({\bf r}'-{\bf R}_j),
\end{equation}
contains the most useful information about the geometric structure of
the cluster. To reduce the number of variables, we rewrite
$\rho_2({\bf r},{\bf r}')$ in terms of the relative
coordinate ${\bf x} ={\bf r}-{\bf r}'$ and the center of mass
coordinate and average over the center of mass coordinate as well as
over the remaining angular degrees of freedom. This defines the
so-called average radial distribution function
$\bar{g}_n (x)$,
\begin{eqnarray}
  \bar{g}_n(x) 
 & =&\: \frac{1}{4\pi x^2 \rho_0n} \sum_{i \neq j}
\delta (x-\left|{\bf R}_i-{\bf R}_j\right|),
 \label{Eq:av_g}
\end{eqnarray}
where $\rho_0$ is the average bulk density.  This procedure yields the
exact pair distribution function for homogeneous systems. Finally, to
obtain a smooth function, we have broadened the distribution of pair
distances $x$ by a Gaussian of width $\epsilon$,
\begin{equation}
  \bar{g}_{n,\epsilon}(x)\equiv
  \frac{1}{4\pi\sqrt{2\pi}\epsilon r^2 \rho_0n}
   \sum_{i \neq j}
   \mathrm{e}^{-\frac{1}{2\epsilon}\left(x-\left|{\bf R}_i-{\bf R}_j\right|\right)^2}\,.
 \label{Eq:av_g_eps}
\end{equation}
For small $\epsilon$, Eq.~\eqref{Eq:av_g_eps} is identical to the
discretized expression in Eq.~\eqref{Eq:av_g}. 

\begin{figure}
  \centering
  \includegraphics[width=\linewidth,keepaspectratio]{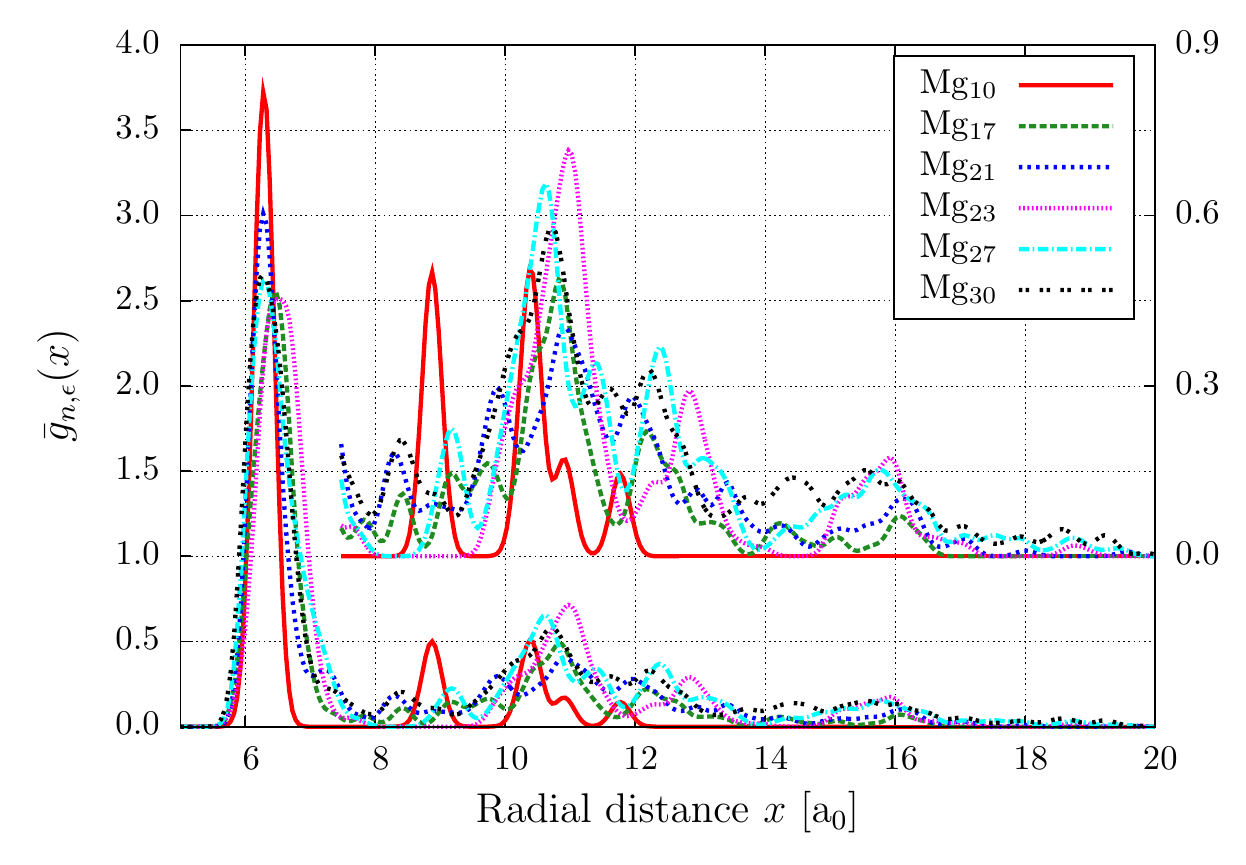}
  \caption{(color online) Pair distribution function for selected Mg$_n$ clusters.
    The inset shows the distribution functions for $ x > 7.5\,a_0$
    magnified (right scale). We have chosen a broadening parameter of
    $\epsilon=0.15\,\mathrm{a_0}$. }
  \label{fig:gr_neut}
\end{figure}

\begin{figure}
  \centering
  \includegraphics[width=\linewidth,keepaspectratio]{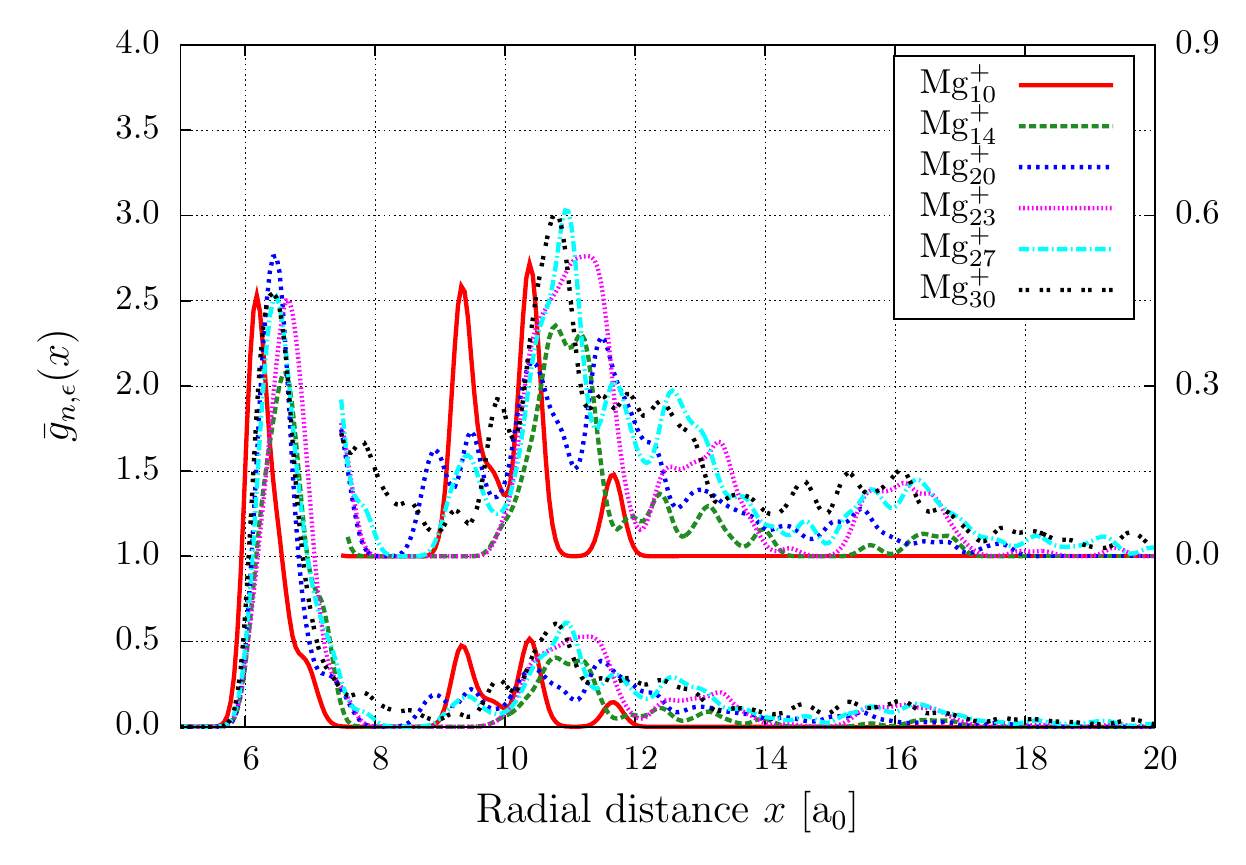}
  \caption{(color online) Pair distribution function for selected
    Mg$_n^+$ clusters. The inset shows the distribution functions for
    $ x > 7.5\,a_0$ magnified (right scale). We have chosen a
    broadening parameter of $\epsilon=0.15\,\mathrm{a_0}$.}
  \label{fig:gr_plus}
\end{figure}

The peaks in $\bar{g}_{n,\epsilon}(x)$ correspond to a
high probability of finding two atoms at a distance $x$ and, in the
bulk limit, correspond to the sequence of $n$-th nearest neighbor
distances $\xi_n$. For face-centered cubic (fcc) and hexagonal close
packed (hcp) structures, the ratios between first and second, and
first and third nearest neighbor distances are
\begin{equation}
  \label{eq:nnfcphcp}
  \xi_2:\xi_1 = \sqrt{2},\;\;
  \xi_3:\xi_1 = \sqrt{3}\qquad \text{(fcc \emph{and} hcp).}
\end{equation}
The pair distribution functions $\bar{g}_{n,\epsilon}(x)$
for a few selected neutral and singly ionized Mg clusters are shown in
Figs.~\ref{fig:gr_neut} and \ref{fig:gr_plus}.
While $\mathrm{Mg}_{30}$ is still far away from the bulk limit, one
can nevertheless clearly identify a sequence of peaks in the plot of
$\bar{g}_{30,\epsilon}(x)$.  In this case, we have
determined the positions of these peaks to be
$\xi_1=6.275\,\mathrm{a_0}$, $\xi_2=8.375\,\mathrm{a_0}$ and
$\xi_3=10.725\,\mathrm{a_0}$, respectively. The corresponding ratios
of peak positions are
\begin{equation}
  \xi_2:\xi_1 = 1.34\, (-5.24\%),\;\; \xi_3:\xi_1=1.71\,(-1.27\%),
\end{equation}
where the values in parentheses give the error relative to the known
ratios for the bulk fcc and hcp lattices as given in
Eq.~\eqref{eq:nnfcphcp}. From our calculations one would therefore
conclude that magnesium clusters approach, with increasing cluster
size, either fcc or hcp structure. The $\mathrm{Mg}_{30}$ cluster is
still too small to reliably extract a fourth peak from
$\bar{g}_{30,\epsilon}(x)$, which would allow to
distinguish between these two structures. However, up to the cluster
sizes taken into account here, our results are consistent with
the known hcp symmetry of bulk magnesium.

We finally draw attention to another remarkable piece of information
that can be gained from Figs.~\ref{fig:gr_neut} and \ref{fig:gr_plus}:
When comparing the average pair distribution functions for the
different clusters in the range from $8\,\mathrm{a_0}$ to
$10\,\mathrm{a_0}$, it is evident that $\bar{g}_{n,\epsilon}(x)$ is
zero for the clusters Mg$_{14}^+$, Mg$_{23}$ and Mg$_{23}^+$, while it
is non-zero for all other clusters examined here. It can be concluded
that those clusters do have a geometric structure that is
qualitatively different from that of the other ``magic'' clusters.

\begin{figure}
  \centering
  \includegraphics[width=\linewidth,keepaspectratio]{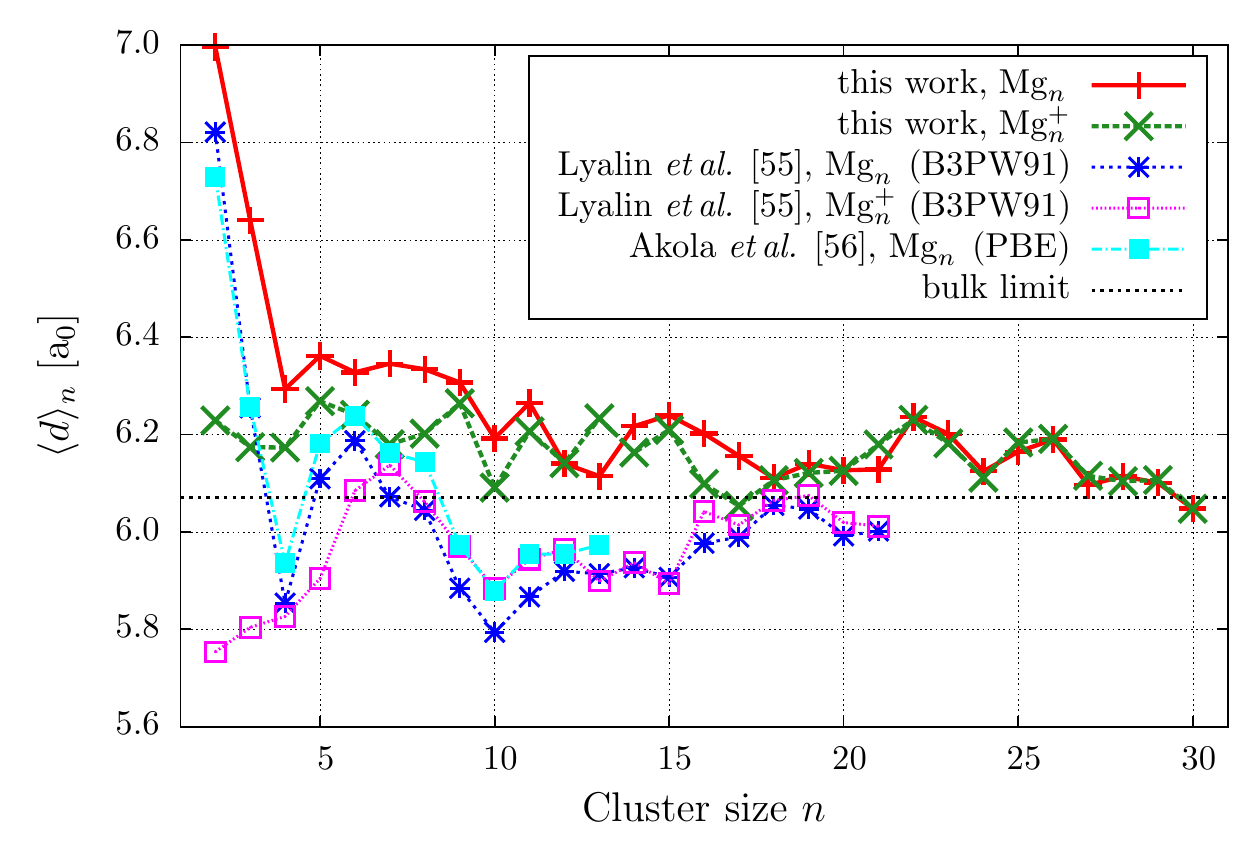}
 \caption{(color online) 
  Average nearest neighbor distance
  $\langle d \rangle_n$ for neutral Mg$_n$ and charged Mg$_n^+$
   clusters as a function of cluster size $n$. Also shown
  is the bulk limit (dotted line) and the theoretical results
  of Refs.~\cite{LSSG03} and \cite{Akola01}.}
  \label{fig:mg-dmean}
\end{figure}

From the geometric configurations of the Mg$_n$ clusters we can
calculate the average nearest neighbor distance $\langle d\rangle_n$
and the smallest and largest nearest neighbor distances
$d_{\mathrm{min},n}$ and $d_{\mathrm{max},n}$, respectively. The
numerical values for these quantities up to a cluster size of $n=11$
are given in Fig.~\ref{fig:mg-clusters}. The evolution of $\langle d
\rangle_n$ with cluster size for both neutral and singly ionized
clusters is shown in Fig.~\ref{fig:mg-dmean}, together with results
from Refs.~\cite{LSSG03} and \cite{Akola01}. The dotted line in this
figure marks the bulk limit of $\langle d \rangle =6.07\,\mathrm{a_0}$
for solid magnesium. Evidently, at $n=30$ both neutral and charged
clusters are already close to that limit.

\subsubsection{Kohn-Sham energies}
\label{sssec:mg-KS}

\begin{figure}
  \centering
  \includegraphics[width=\linewidth,keepaspectratio]{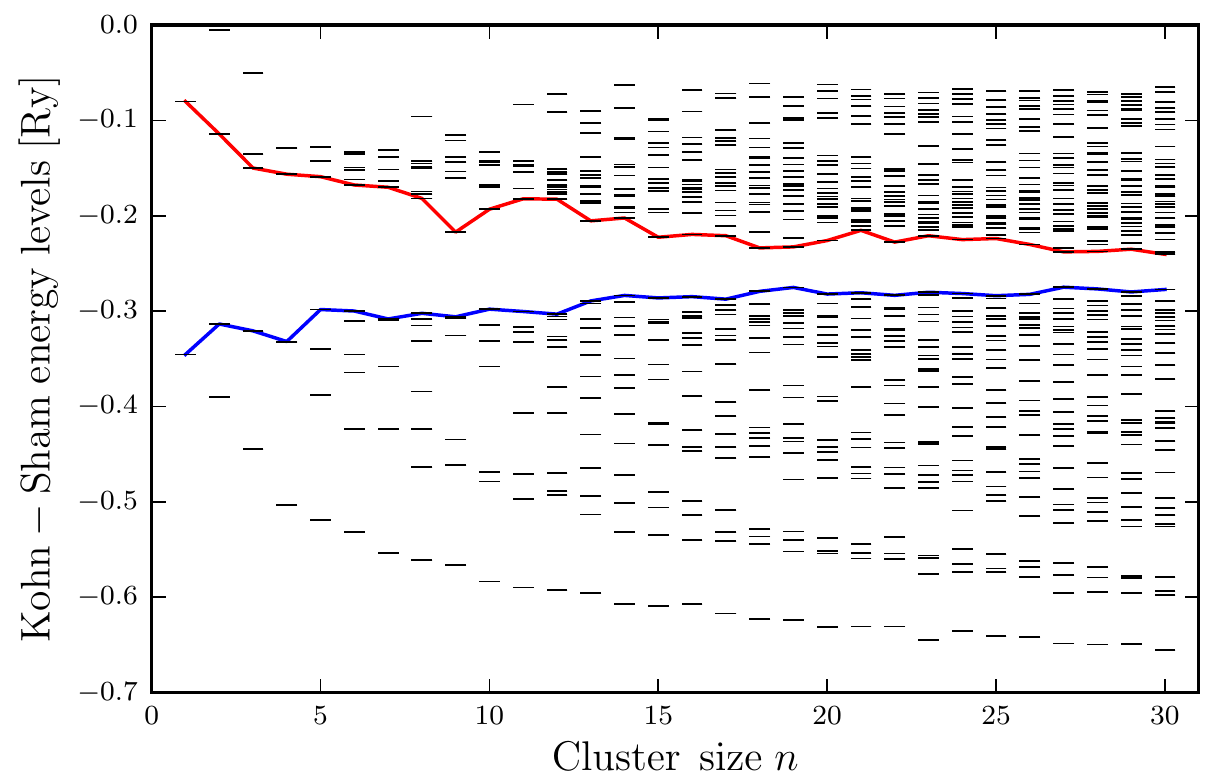}
  \caption{(color online) Kohn-Sham energy levels of the neutral Mg$_n$ clusters as a
  function of cluster size. The HOMO levels are marked by the lower (blue)
  solid line, the LUMO levels by the upper (red) solid line.}
  \label{fig:mg-KS_levels_neut}
\end{figure}

\begin{figure}
  \centering
  \includegraphics[width=\linewidth,keepaspectratio]{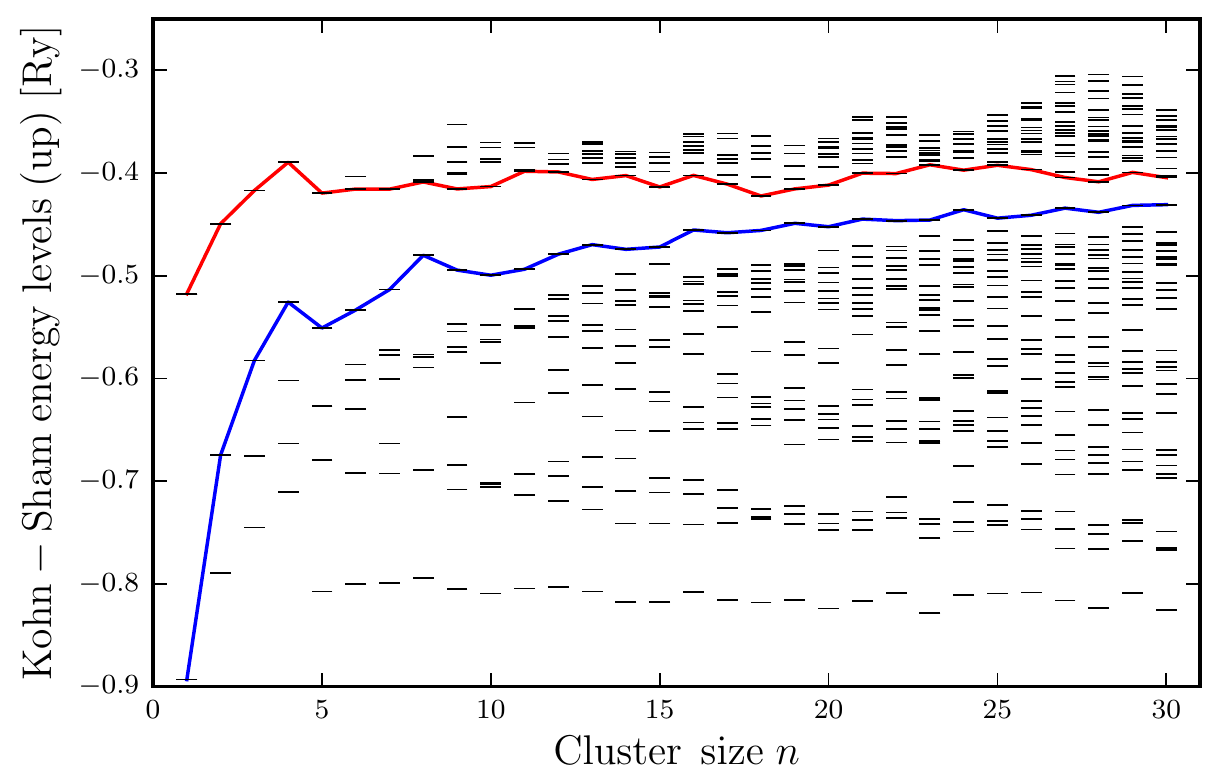}
  \includegraphics[width=\linewidth,keepaspectratio]{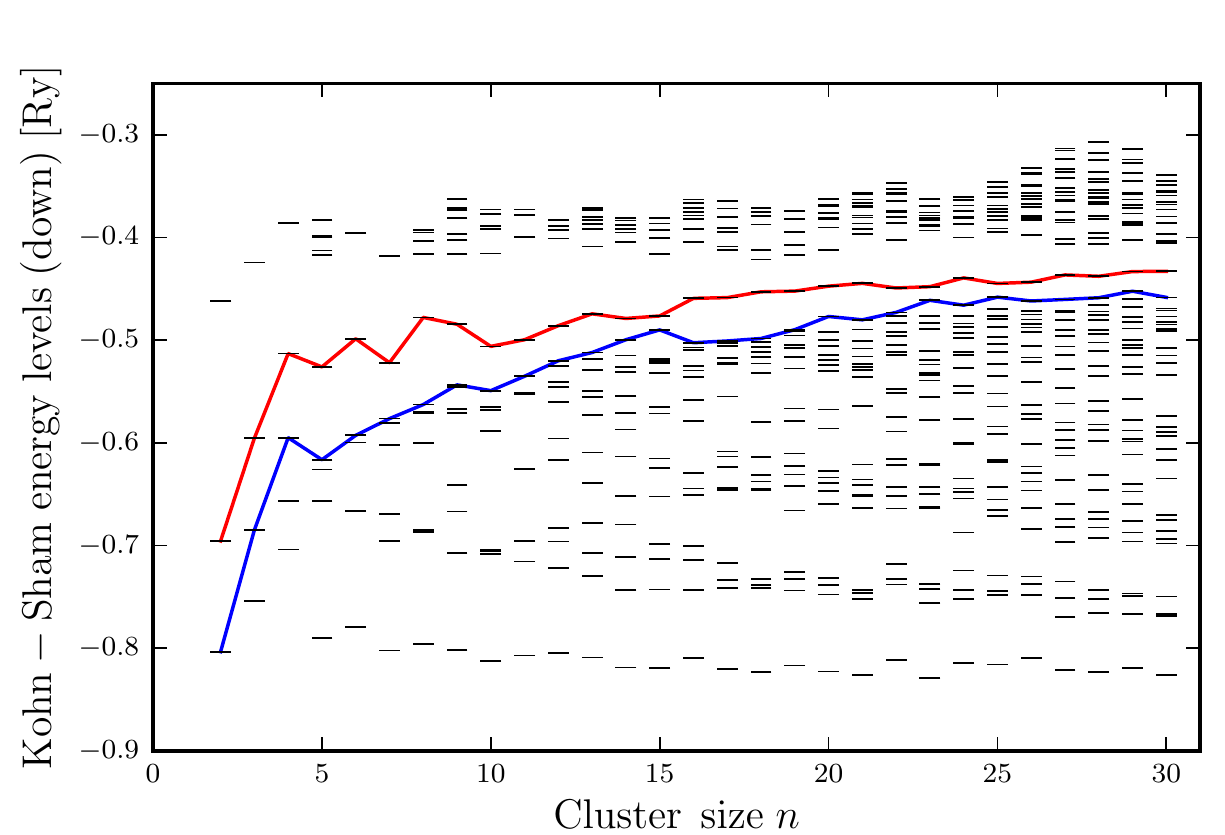}
  \caption{(color online) Kohn-Sham energy levels of the singly ionized Mg$^+_n$ clusters as a
  function of cluster size. Top pane: Energy levels for spin-up
  electrons. Bottom pane: Energy levels for spin-down
  electrons. }
  \label{fig:mg-KS_levels_plus}
\end{figure}

The Kohn-Sham levels of the ground state structures of Mg$_n$
and Mg$_n^+$ are shown in Figs.~\ref{fig:mg-KS_levels_neut} and
\ref{fig:mg-KS_levels_plus}. For the Mg$_n^+$ clusters, the LSDA has
been used; we therefore show the energy levels for both spin channels
in separate plots.

In contrast to the HOMO-LUMO gaps of the sodium clusters shown in
Figs.~\ref{fig:na-KS_levels_neut} and \ref{fig:na-KS_levels_plus}, the
gaps of the magnesium clusters do not show strong signs of electronic
shell closures: The gap decreases rather monotonically with cluster
size; the pronounced ``zigzag'' structure of the HOMO and LUMO levels
found in sodium is missing. This is an indication that small Mg
clusters are more covalently bound and electrons are more localized in
comparison to sodium clusters. This prediction is confirmed by the
respective electron densities, see Sec.~\ref{sec:mgnarho}.

In a bulk system, the closing of the HOMO-LUMO gap can be considered
as a sign of an NMM transition. In finite-sized systems like clusters,
the gap between energy levels is always finite, and the answer to the
question whether a material is a metal or not depends on its
temperature \cite{IssChe05}. Specifically, a temperature of 300 Kelvin
corresponds to roughly 1.9$\cdot10^{-3}\,$Ry. The calculated HOMO-LUMO
gaps shown in Figs.~\ref{fig:mg-KS_levels_neut} and
\ref{fig:mg-KS_levels_plus} are on the order of 0.02--0.10~Ry. Therefore,
Mg clusters with up to 30 atoms are not metallic by this
criterion. Note that the LDA is known to systematically
\emph{underestimate\/} band gaps (see, e.g.,
Ref. \cite{vanschilfgaarde2006}), and is thus erring on the side of
caution when predicting non-metallicity.

At first glance, this conflicts with
Refs. \cite{DDB01,diederichPRA05,Thomas02}, where the NMM
transition has been observed for Mg clusters in helium at around
$n=20$. However, Refs. \cite{Jellinek02,Acioli213402} suggest that the
observed NMM transition most likely refers to {\em negatively charged}
Mg$_n^-$ clusters used in these experiments and thus does not
disagree with our findings.

\subsubsection{Electron density}
\label{sec:mgnarho}

\begin{figure}
  \centering
  \includegraphics[width=0.9\linewidth,keepaspectratio]%
  {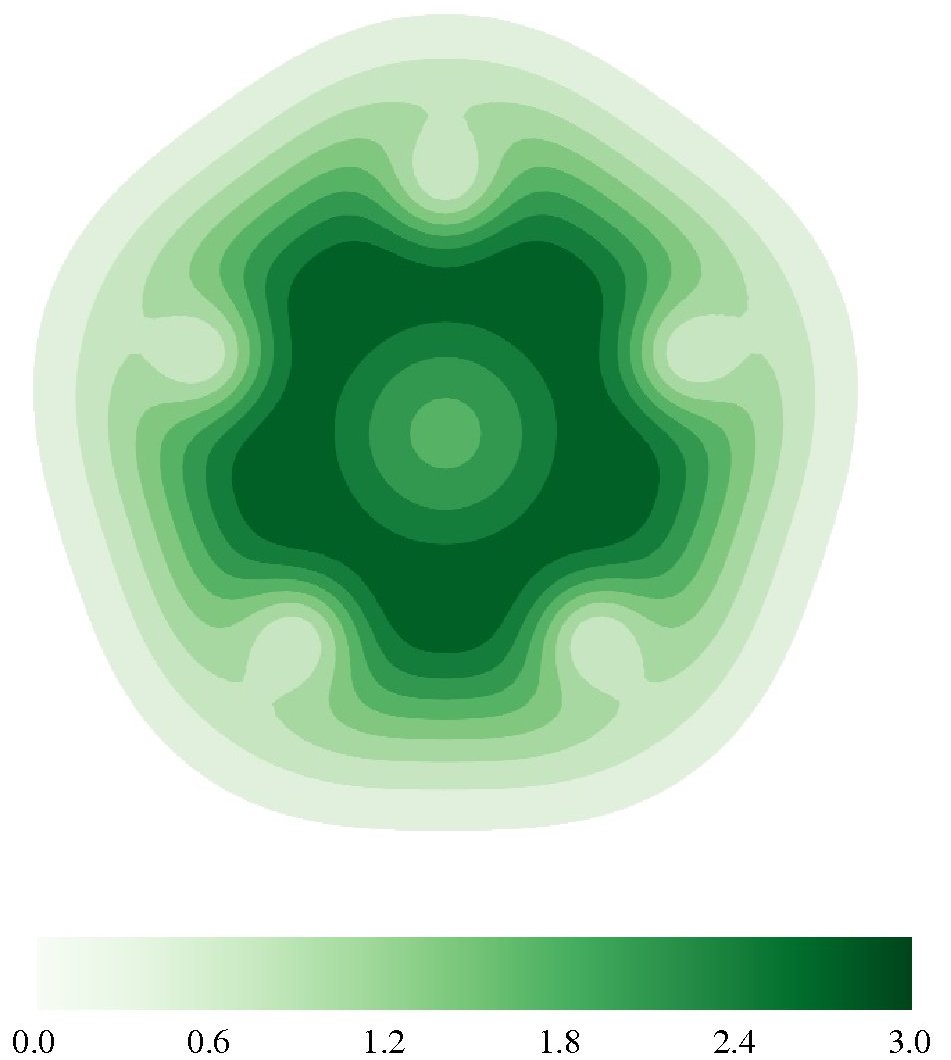}
  \caption{(color online) Contour plot of the electron density of a Na$_7$ cluster in the
  symmetry plane, in units of $10^{-3}\,\mathrm{a_0}^{-3}$.}
  \label{fig:na-edens_07}
\end{figure}

\begin{figure}
  \centering
  \includegraphics[width=0.9\linewidth,keepaspectratio]%
  {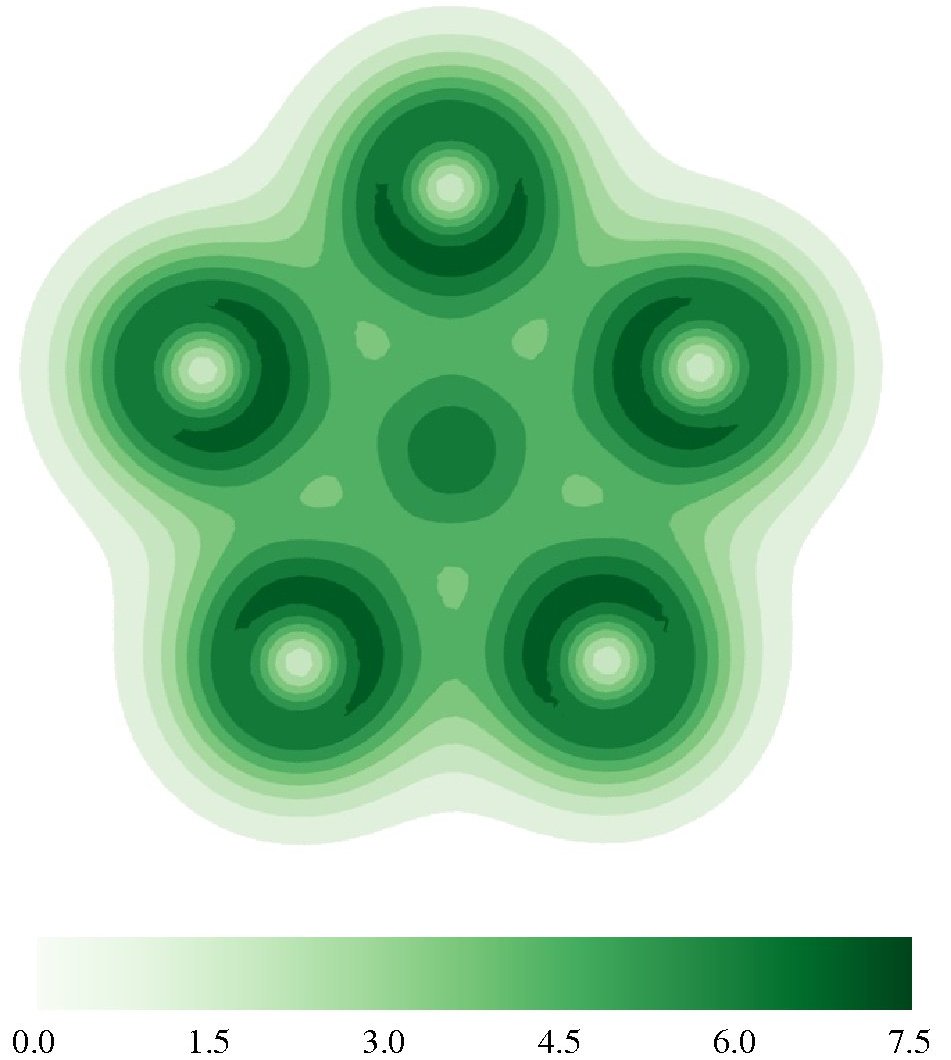}
  \caption{(color online) Contour plot of the electron density of a Mg$_7$ cluster in the
  symmetry plane, in units of $10^{-3}\,\mathrm{a_0}^{-3}$.}
  \label{fig:mg-edens_07}
\end{figure}

The most telling documentation of the difference between Na$_n$ and
Mg$_n$ clusters is the electron density. In Figs.
\ref{fig:na-edens_07} and \ref{fig:mg-edens_07} we show, as examples,
contour plots of the electron density of the Na$_7$ and Mg$_7$ ground
state configuration in the symmetry plane. We have chosen these
examples because their ionic configuration is very similar, see
Figs.~\ref{fig:na-clusters} and \ref{fig:mg-clusters}.


\begin{figure}
  \centering
  \includegraphics[width=0.9\linewidth,keepaspectratio]%
  {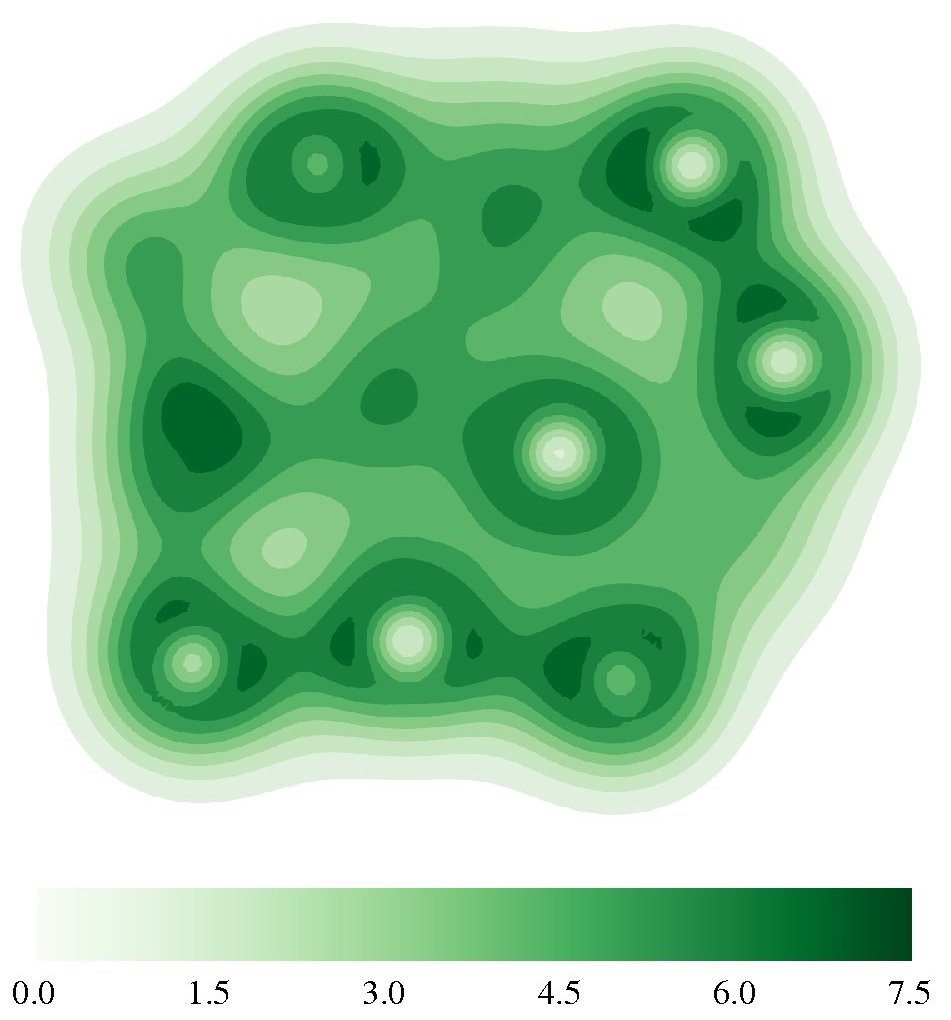}
  \caption{(color online) Contour plot of the electron density of a
Mg$_{30}$ cluster in the
  $yz$-plane, in units of $10^{-3}\,\mathrm{a_0}^{-3}$.}
  \label{fig:mg-edens_30}
\end{figure}

From Figs.~\ref{fig:na-edens_07} and \ref{fig:mg-edens_07} it is clear
that the electrons in a Na$_7$ cluster are much more delocalized than
in a Mg$_7$ cluster. If there were an NMM transition in
the range of cluster sizes under consideration here, one would expect
a delocalization of electrons and a much broader electron density for
larger clusters. In Fig.~\ref{fig:mg-edens_30} we show therefore a
contour plot of the electron density in a Mg$_{30}$ cluster. Note that
the scales are the same as in Fig.~\ref{fig:mg-edens_07}. Evidently,
the electrons are just as strongly localized as in Mg$_7$, indicating
that {\em neutral\/} clusters have, at that ion number, not yet
undergone an NMM transition.

\subsubsection{Isomers}
\label{sec:isomers}

Tables \ref{tab:mg-isomers_neut} and \ref{tab:mg-isomers_plus} show
the binding energies per atom, $E_{\mathrm{b},n}^{(+)}$, for the
ground states of neutral
Mg$_n$ and singly charged Mg$_n^+$ clusters, together with the
\emph{differences} in binding energy per atom of identified isomers to
the corresponding ground state.

\begin{table}
 \centering
 \begin{tabular}{l r@{.}l r@{.}l r@{.}l r@{.}l}
  \toprule
  Mg$_n^{\phantom{+}}$ & \multicolumn{2}{c}{$E_{\mathrm{b},n}^{\phantom{+}}/n$[Ry]} & \multicolumn{2}{c}{$b\,$[10$^{-3}$Ry]} & \multicolumn{2}{c}{$c\,$[10$^{-3}$Ry]} & \multicolumn{2}{c}{$d\,$[10$^{-3}$Ry]}\\
  \midrule
  Mg$_2^{\phantom{+}}$    & 0&004904 \\
  Mg$_3^{\phantom{+}}$    & 0&011110 & -3&653 \\
  Mg$_4^{\phantom{+}}$    & 0&020072 \\
  Mg$_5^{\phantom{+}}$    & 0&021798 \\
  Mg$_6^{\phantom{+}}$    & 0&023444 & -0&291 \\
  Mg$_7^{\phantom{+}}$    & 0&026227 & -0&450 & -0&606 \\
  Mg$_8^{\phantom{+}}$    & \phantom{--}0&028047 & \phantom{--}-0&278 & \phantom{--}-0&798 & \phantom{--}-1&215 \\
  Mg$_9^{\phantom{+}}$    & 0&031079 & -1&360 \\
  Mg$_{10}^{\phantom{+}}$ & 0&033874 & -0&357 & -1&827 \\
  Mg$_{11}^{\phantom{+}}$ & 0&034759 & -1&572 & -2&004 \\
  Mg$_{12}^{\phantom{+}}$ & 0&034859 & -0&547 & -0&762 \\
  Mg$_{13}^{\phantom{+}}$ & 0&035357 & -0&226 & -0&447 & -1&219 \\
  Mg$_{14}^{\phantom{+}}$ & 0&037420 & -0&563 & -0&590 & -0&889 \\
  Mg$_{15}^{\phantom{+}}$ & 0&039016 & -0&869 \\
  \bottomrule
 \end{tabular}
 \caption{Binding energies per atom, $E_{\mathrm{b},n}/n$, of the ground state configurations
   and a few isomers of neutral Mg$_n$ clusters up to $n = 15$. The
   second column shows $E_{\mathrm{b},n}/n$ for the
   ground-state configuration, columns (b)-(d) show the
   {\em difference in binding energy per atom\/} between the ground state
   and the second, third, and fourth isomer.}
 \label{tab:mg-isomers_neut}
\end{table}
\begin{table}
 \centering
 \begin{tabular}{l r@{.}l r@{.}l r@{.}l r@{.}l}
  \toprule
  Mg$_n^+$ & \multicolumn{2}{c}{$E_{\mathrm{b},n}^+/n$[Ry]} & \multicolumn{2}{c}{$b\,$[10$^{-3}$Ry]} & \multicolumn{2}{c}{$c\,$[10$^{-3}$Ry]} & \multicolumn{2}{c}{$d\,$[10$^{-3}$Ry]}\\
  \midrule
  Mg$_2^+$    & 0&048724 \\
  Mg$_3^+$    & 0&053346 \\
  Mg$_4^+$    & 0&052536 & -1&382 \\
  Mg$_5^+$    & 0&053102 & -0&902 \\
  Mg$_6^+$    & \phantom{--}0&052759 & \phantom{--}-1&296 & \phantom{--}-1&622 & \phantom{--}+2&235 \\
  Mg$_7^+$    & 0&051013 & -0&013 & -0&182 \\
  Mg$_8^+$    & 0&051870 & -0&008 & -0&892 \\
  Mg$_9^+$    & 0&051520 & -0&024 & -0&214 \\
  Mg$_{10}^+$ & 0&053659 & -0&049 & -0&485 \\
  Mg$_{11}^+$ & 0&053859 & -0&333 & -1&496 \\
  Mg$_{12}^+$ & 0&053125 & -0&304 & -1&051 \\
  Mg$_{13}^+$ & 0&053817 & -0&184 & -0&379 & -0&425 \\
  Mg$_{14}^+$ & 0&054982 & -0&335 & -0&375 & -0&577 \\
  Mg$_{15}^+$ & 0&055399 & -0&268 \\
  \bottomrule
 \end{tabular}
 \caption{Binding energies per atom, $E_{\mathrm{b},n}^+/n$, of the ground state configurations
   and a few isomers of charged Mg$_n^+$ clusters up to $n = 15$. The
   second column shows $E_{\mathrm{b},n}^+/n$ for the
   ground-state configuration, columns (b)-(d) show the
   {\em difference in binding energy per atom\/} between the ground state
   and the second, third, and fourth isomer.}
 \label{tab:mg-isomers_plus}
\end{table}

\begin{figure}
  \centering
  \includegraphics[width=0.45\linewidth,keepaspectratio]{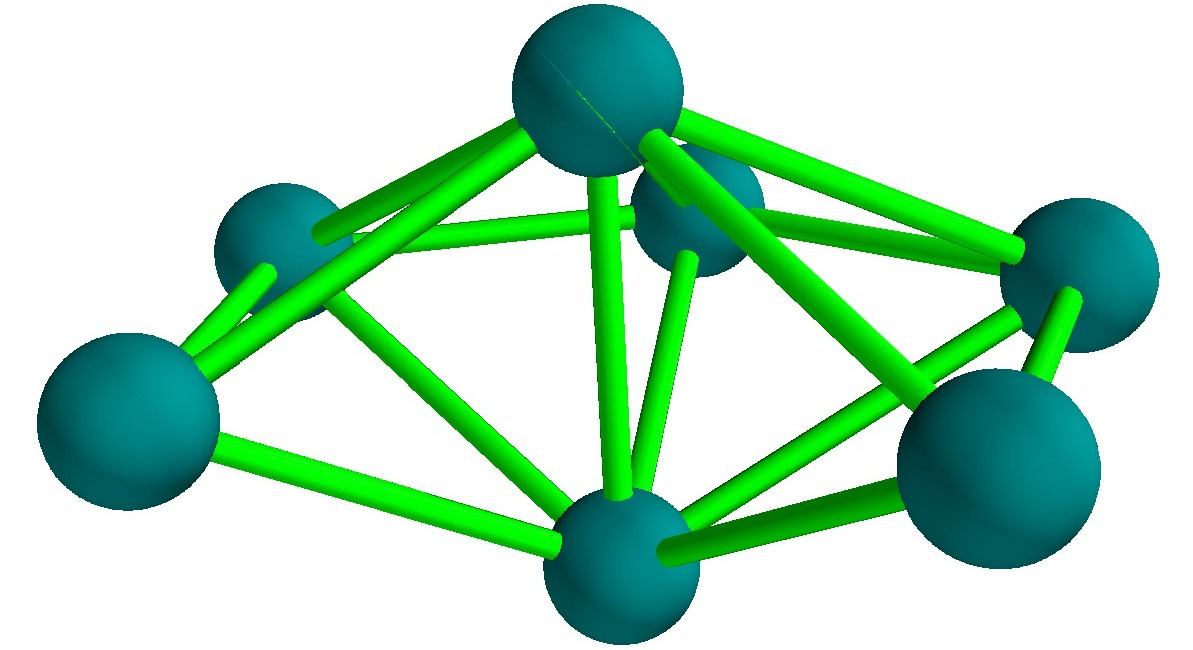}
  \hspace{0.05\linewidth}
  \includegraphics[width=0.45\linewidth,keepaspectratio]{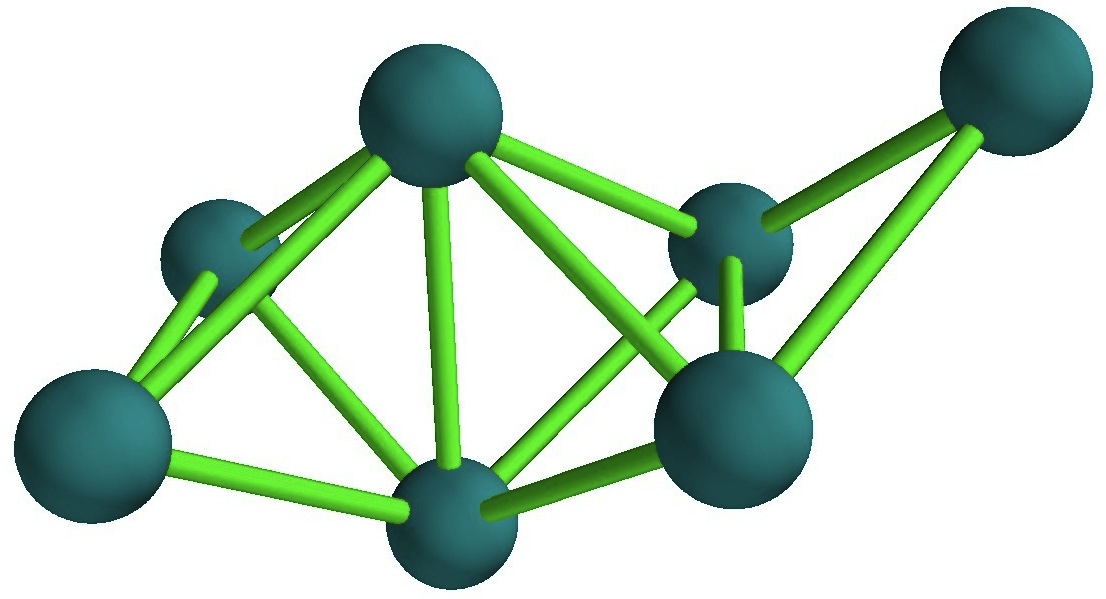} 
 \caption{(color online) The two structures of Mg$_7^+$ with lowest
    energy. Left pane: ground state configuration. Right pane: lowest-lying
    isomer (energy difference: $-1.3\cdot10^{-5}\,\mathrm{Ry}$, see
    Tab.~\ref{tab:mg-isomers_plus}).}
  \label{fig:mg-compare_07+}
\end{figure}

\begin{figure}
  \centering
  \includegraphics[width=0.45\linewidth,keepaspectratio]{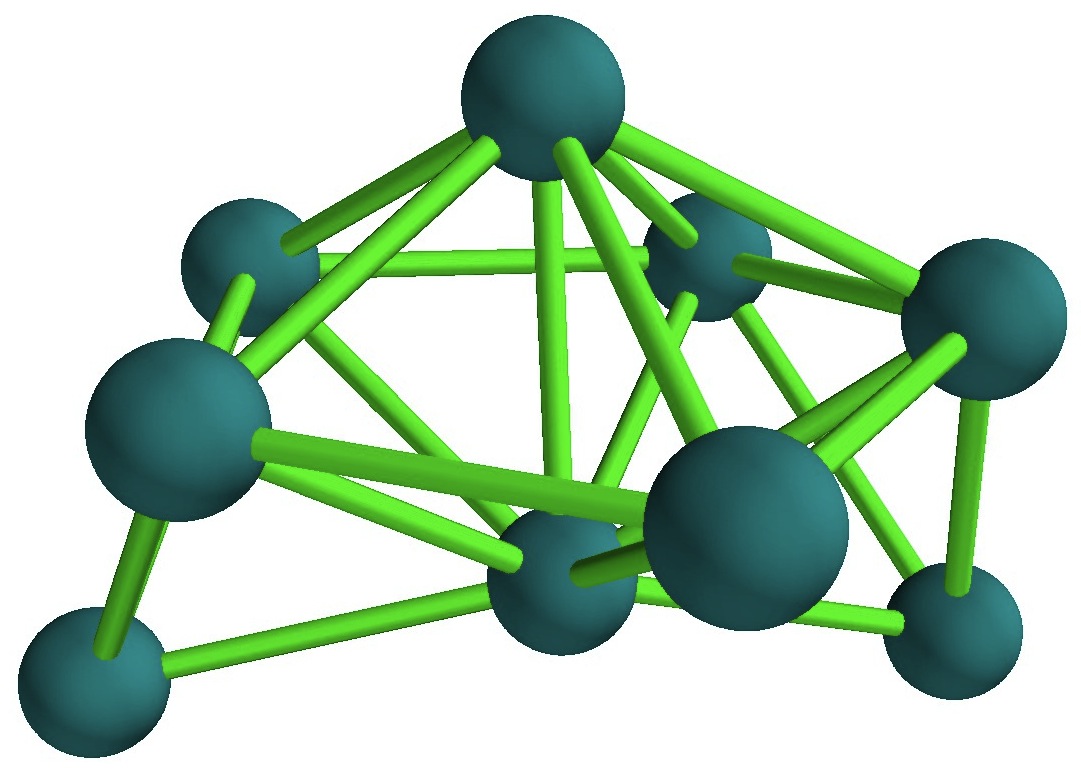}
  \hspace{0.05\linewidth}
  \includegraphics[width=0.45\linewidth,keepaspectratio]{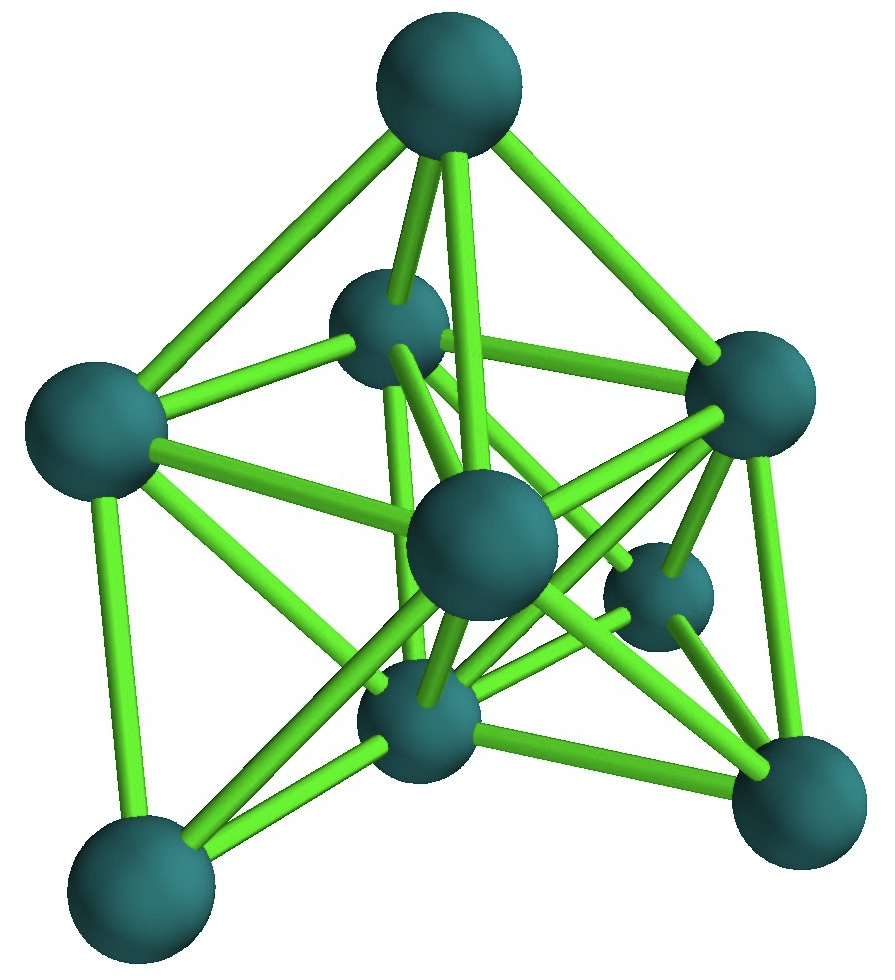}
  \caption{(color online) The two structures of Mg$_9^+$ with lowest
    energy. Left pane: ground state configuration. Right pane: lowest-lying
    isomer (energy difference: $-2.4\cdot10^{-5}\,\mathrm{Ry}$, see Tab.~\ref{tab:mg-isomers_plus}).}
  \label{fig:mg-compare_09+}
\end{figure}

A remarkable feature of these results can be seen in Table
\ref{tab:mg-isomers_plus}: For all clusters between Mg$_7^+$ and
Mg$_{10}^+$, the lowest-lying isomer is energetically almost
degenerate with the ground state. Figs.~\ref{fig:mg-compare_07+} and
\ref{fig:mg-compare_09+} show the corresponding geometric structures
for Mg$_7^+$ and Mg$_9^+$, respectively.  In the case of Mg$_7^+$, see
Fig.~\ref{fig:mg-compare_07+}, both the ground-state and the
lowest-lying isomer are composed of the same Mg$_6^+$ core structure,
with the additional atom added at two different positions. For
Mg$_9^+$, shown in Fig.~\ref{fig:mg-compare_09+}, the situation seems
to be more complicated. Here the atom positions and bonding angles for
the two configurations are considerably different.

\section{Conclusion}
\label{sec:conclusion}

We have presented a systematic study of neutral and singly charged
Na$_n$ and Mg$_n$ clusters. We have employed a real-space algorithm
that is particularly well suited for the problem at hand: It allows a
fine coordinate-space resolution that is able to deal with even small
annealing steps. The process is iterative and profits from the
knowledge of accurate solutions of near-by configurations. We found,
during the annealing process, that the HOMO-LUMO gaps are indeed very
sensitive to the details of the ground state structure. Thus, the last
annealing steps were of the order of 10$^{-4}$\,a$_0$. We have
verified the numerical stability of our procedure by moving and
rotating the cluster with respect to the coordinate space mesh.  The
{\em total\/} energy had uncertainties comparable to what is gained
during the final annealing steps, but the relative energies between
neighboring configurations were independent of the precise location
and orientation of the cluster.

The relatively strongly localized electron density in Mg clusters also
suggests that a simple jellium model can not describe the cluster
properties accurately. In contrast to alkali clusters in general,
where the delocalized electrons dominate the cluster structure, the
geometric symmetry seems to be more important for Mg clusters. The
highly symmetric Mg$_{11}$ cluster has, for instance, only a slightly
higher binding energy per atom than Mg$_{10}$, for which the
jellium model predicts an electronic shell closure. Nevertheless, the
stability analysis depicted in Fig.~\ref{fig:mg-deltae_neut} shows that
Mg$_{10}$ is still more stable than Mg$_{11}$.  This is different for
Mg$_{20}$ and Mg$_{21}$: The jellium model predicts a shell closure
for Mg$_{20}$, but our calculations show that Mg$_{21}$ is slightly
more stable than Mg$_{20}$.

Because Mg$_n$ and Mg$_n^+$ clusters with up to 30 do not show closing
HOMO-LUMO gaps, we conclude that these clusters are not metallic. This
is consistent with the findings of
Refs.~\cite{Jellinek02,Acioli213402} that the observed NMM transition
for Mg clusters in helium at around $n=20$
\cite{DDB01,diederichPRA05,Thomas02} most likely refers to the {\em
  negatively charged} Mg$_n^-$ clusters used in the corresponding
experiments.

\bibliography{papers}
\bibliographystyle{epj}

\end{document}